\documentclass{aa}
\usepackage{graphicx}
\usepackage{txfonts}
\usepackage{diagbox}
\usepackage{rotating}
\usepackage{tikz}
\usepackage{multirow}
\usepackage{wasysym}
\usepackage{adjustbox}
\usepackage{afterpage}
\usepackage{placeins}
\usepackage{xcolor}
\usepackage{caption}
\usepackage[strict]{changepage}
\usepackage{lscape}
\usepackage{pdflscape}
\usepackage{soul}

\usepackage{hyperref}
\hypersetup{colorlinks=True,citecolor=blue}

\def\ms{\hbox{\,m\,s$^{-1}$} }         %m.s -1
\def\m2s2{\hbox{\,m$^{2}$\,s$^{-2}$} } %m2.s -2
\def\ang{\text{\AA}}

\usepackage{amstext}

\begin{document}

   \title{YARARA: Significant improvement of RV precision through post-processing of spectral time-series}

  % \subtitle{I. Description and performance of the code}

   \author{M. Cretignier
          \and X. Dumusque 
          \and N. C. Hara 
          \and F. Pepe 
          }

   \institute{Astronomy Department of the University of Geneva, 51 ch. des Maillettes, 1290 Versoix, Switzerland\\
              \email{michael.cretignier@unige.ch}}

   \date{Received XXX ; accepted XXX}

% \abstract{}{}{}{}{} 
% 5 {} token are mandatory
 
  \abstract
  % context heading (optional)
  % {} leave it empty if necessary  
   {}
  % aims heading (mandatory)
   {Even the most-precise radial-velocity instruments gather high-resolution spectra that present systematic errors that a data reduction pipeline cannot identify and correct for efficiently by analysing a set of calibrations and a single science frame. In this paper, we aim at improving the radial-velocity precision of HARPS measurements by 'cleaning' individual extracted spectra using the wealth of information contained in spectra time-series.}
  % methods heading (mandatory)
   {We developed YARARA, a post-processing pipeline designed to clean high-resolution spectra from instrumental systematics and atmospheric contamination. Spectra are corrected for: tellurics, interference pattern, detector stitching, ghosts and fiber B contaminations as well as more advanced spectral line-by-line corrections. YARARA uses Principal Component Analysis on spectra time-series with prior information to disentangle contaminations from real Doppler shifts. We applied YARARA on three systems: HD10700, HD215152 and HD10180 and compared our results to the HARPS standard Data Reduction Software and the SERVAL post-processing pipeline.}
  % results heading (mandatory)
   {We run YARARA on the radial-velocity data set of three stars intensively observed with HARPS: HD10700, HD215152 and HD10180. On HD10700, we show that YARARA enables to obtain radial-velocity measurements that present a rms smaller than 1\,\ms over the 13 years of the HARPS observations, which is 20 and 10 \% better than the HARPS Data Reduction Software and the SERVAL post-processing pipeline, respectively. We also injected simulated planets on the data of HD10700 and demonstrated that YARARA does not alter pure Doppler shifted signals. On HD215152, we demonstrated that the 1-year signal visible in the periodogram becomes marginal after processing with YARARA and that the signals of the known planets become more significant. Finally, on HD10180, the known six exoplanets are well recovered although different orbitals parameters and planetary masses are provided by the new reduced spectra.}  
   {Post-processing correction of spectra using spectra time-series allows to significantly improve the radial-velocity precision of HARPS data and demonstrate that for the extremely quiet star HD10700, a radial-velocity root mean square better than 1 m/s can be reached over the 13 years of HARPS observations. Since the processing proposed in this paper does not absorb planetary signals, its application to system intensively followed is promising and will certainly push further the detection of the lightest exoplanets.}


   \maketitle

\section{Introduction}

During the last 25 years, the discoveries of exoplanets by radial velocity (RV) has considerably improved our knowledge about the diversity of exoplanetary systems. Nevertheless, detection of exoplanets is somewhat biased toward the detection of massive and short-periods planets. The discovery of a sub meter-per-second planetary signal with a period comparable or longer than a year remains extremely challenging due to several factors such as instrumental systematics, atmospheric contamination and stellar activity. 

Obtaining extremely precise radial velocities (EPRV) is an art that requires 1) a high-resolution super-stable spectrograph with a high spectral fidelity to provide the cleanest possible spectra at the pre-calibration stage, 2) a data reduction pipeline that extract at best the spectra from raw frames, correct the instrumental systematics using a set of calibration frames and extract relevant information like the final radial-velocity of the spectrum and 3) a post-processing of the RV time-series to correct for stellar activity signals.

Due to tremendous progresses in instrumentation, point 1) is becoming less of a concern due to the new generation of EPRV spectrographs such as ESPRESSO \citep{Pepe(2014),Pepe(2021)}, EXPRES \citep{Fischer(2017)} and MAROON-X \citep{Seifahrt(2018),Peck(2020),Sanchez(2021)} that already demonstrated RV measurements with root mean square (rms) significantly smaller than 1\,\ms. 
The correction of stellar activity in the RV time-series, point 3) above, has also been widely studied \citep[e.g.][]{Haywood(2014),Grunblatt(2015), Davis(2017), Jones(2017), Dumusque(2018), Cretignier(2020a), Kosiarek(2020),Collier(2020)}. There is only point 2) above, corresponding to a better extraction of the radial-velocity content of a spectra that was not intensively studied during the last decade. Only a few papers addressed this question recently \citep{Dumusque(2018),Errmann(2020),Zechmeister(2020), Trifonov(2020)}. However, this step is essential to better understand stellar activity if an improvement in RV precision can be obtained.

It is known that different systematics on the spectra introduce RV effects. As an example, we can cite the color variation of the spectrum with airmass or change in atmospheric conditions, tellurics and micro-tellurics \citep{Artigau(2014)}, but also more instrumental-related systematics specific to the HARPS instrument, such as the detector stitching \citep{Dumusque(2015)}. All these contaminations can somewhat be tricky to resolve when considering individual spectra and corrections can only be achieved by successfully modelling the effects using prior information. 

A methodology to correct for these features is to consider spectra time-series, instead of individual spectra, as the former contains a richness of information far more extended than the latter. For example, this methodology show its potential in the WOBBLE code \citep{Bedell(2019)}, designed to remove telluric lines by performing a principal component analysis (PCA) or in the Gaussian process (GP) framework developped by \citet{Rajpaul(2020)}.

Following the WOBBLE approach, we developed a post-processing radial-velocity pipeline called YARARA to cope with any known systematics in HARPS spectra. We will show that by using the power of PCA on HARPS spectra time-series, we can correct for any known systematics without the need of models for the different effects and with a limited amount of prior information. Such a data-driven approach is well suited to be applied on high-signal-to-noise ratio (S/N) spectra of ultra-stable spectrographs, and on stars that have been intensively observed. Fortunately, such data set are the only ones on which we can hope to search for sub-meter-per-second planetary signals, thus the methodology used by YARARA is perfectly fitted to reach the best RV precision and look for the lightest exoplanets.

In this paper, we apply our new post-processing pipeline YARARA on HARPS  \citep{Pepe(2002b), Mayor(2003), Pepe(2003)} data to demonstrate that further improvements can still be accomplished on HARPS RV measurements. The theory behind YARARA is described in Sect.~\ref{sec:theory}. The different corrections performed in this work are described in the same order than the way they are implemented in the pipeline for HARPS, namely by correcting: cosmic rays (Sect.~\ref{sec:cosmics}), interference pattern (Sect.~\ref{sec:fringing}), telluric lines (Sect.~\ref{sec:telluric}), stellar lines variations (Sect.~\ref{sec:activity}), ghosts (Sect.~\ref{sec:ghost}), stitchings (Sect.~\ref{sec:stitching}), simultaneous calibration contamination (Sect.~\ref{sec:thar}) and continuum correction (Sect.~\ref{sec:smooth}). The line-by-line (LBL) RV derivation is described in \ref{sec:mask} and a more advanced LBL correction is presented in Sect.~\ref{sec:lbl}. The applications of YARARA on three stellar systems is presented in Sect.~\ref{sec:results} (respectively Sect.~\ref{sec:hd10700} for HD10700, Sect.~\ref{sec:hd215152} for HD215152 and Sect.~\ref{sec:hd10180} for HD10180). We then conclude in Sect.~\ref{Conclusion}.

\section{Description of the pipeline}\label{sec:theory}

The pipeline can be described as a sequential cascade of recipes applied to correct for the different systematics. The sequence described below is optimized for HARPS and was chosen based on two criteria: i) the constrain we can have on the correction and ii) the amplitude of the systematic in flux. Having a correction that is well constrained allows us to reduce the degeneracy with other recipes. Moreover, since we mainly use multi-linear regression to perform our corrections, it is necessary to correct first for the contaminations with the largest flux variance. 

We note that different instruments might have a different sequential cascade depending on if the same or other contaminations exist and if they have different intensities.
%\st{and is different for each instrument depending on the intensity of the contaminations and if some of them can be neglected or not. The order of the sequence was decided based on criteria. due to the data-driven nature of our pipeline. First, by how much the problem is constrained, and second, is the contamination producing strong flux variations. These two simple mantra are sufficient to edict a good list of corrections which ensures the pipeline to converge to the good solution. If the problem is well constrained, degeneracies with the other recipes are reduced. Moreover, since multi-linear regression is our main tool, it is important to correct in priority the  contaminations with the largest flux variance.}

The assumption behind the sequential approach is to assume that the different contaminations are orthogonal to each other. We will see further that this is the case if prior information (e.g. systematic structure, wavelength domain) is used to constrain the different recipes and if the baseline of the observations is long enough to avoid spurious correlations. As a consequence, YARARA is mainly applicable to system intensively observed and for which a high S/N (>100) can be obtained. 
%\st{Those systems being the ones which can lead to detect exoplanets with semi-amplitude smaller than 1 m/s and period longer than several months. Another kind aspect by regards to HARPS systematics, is that most contaminations are leaving in a specific and known wavelength domain which ensures even more orthogonality between the recipes.}

Our approach also assumes that the point-spread function (PSF) does not vary significantly over time, and thus the data must come from the same stable spectrograph. We note that on HARPS, the change of the optical fibers in 2015 \citep{LoCurto(2015)} significantly changed the instrumental PSF. We therefore have to consider two instruments, one before and one after the upgrade. However, those two instruments are suffering from the same systematics, hence the same cascade of recipes can be applied. The only restriction is that spectra time-series objects before and after the fiber upgrade cannot be mixed in the pipeline and must be reduced separately.

In this work we will call a river diagram map, the map representing in each row a spectrum residual with respect to some reference. We note that we will represent river diagrams always as the difference with respect to the median spectrum of the spectra time-series. %As a remark, the exact criterion to add a planet in the fit also depends on the S/N of the observations and the instrumental resolution. %For HARPS spectra, YARARA is adding in the pre-fitted model every signal with a semi-amplitude larger than $7$ m/s and an eccentricity $e$ smaller than $0.85$. 
Since all the spectra are on a common wavelength grid, the spectra time-series can be seen as a matrix. We will refer to the variation of the flux $f(\lambda_j,t)$ as a function of the time $t$ for a specific wavelength $\lambda_j$ in this matrix as "wavelength column". Each column is assumed independent for simplicity. 
The general mathematical framework of the corrections applied in YARARA take the shape of a multi-linear model fitted on the wavelength columns: 

\begin{equation}
\label{eq1}
\delta f(\lambda_j,t) = cst + \sum_{i=1}^{N} \alpha_i(\lambda_j) \cdot P_i(t),
\end{equation}

where $\alpha_i(\lambda_j)$ are the strength coefficients, fitted for using a weighted least-square considering as weight the inverse of the flux uncertainty squared, $P_i(t)$ are  the elements of the basis fitted and $\lambda_j$ are the wavelengths either in the stellar or terrestrial rest-frame. The data $\delta f(\lambda_j,t)$ can be either spectra difference or ratio by respect to a master spectrum $f_0(\lambda_j)$ depending on the systematics that we want to correct for. Each time a RV shift of the spectra is performed, a cubic interpolation is used to resample the shifted spectra on a common $\lambda_j$ wavelength grid. The $P_i(t)$ basis will be made either of CCFs moments (contrast and FWHM) obtained by a weighted least-square Gaussian fit, or principal components of a weighted PCA \citep{Bailey(2012)} trained on a specific set of wavelengths $\lambda_j$. In all cases, the weights are defined as the inverse of the flux uncertainty squared.

\subsection{Data preprocessing  \label{sec:preprocess}}

\begin{figure*}[h]
	\includegraphics[width=18cm]{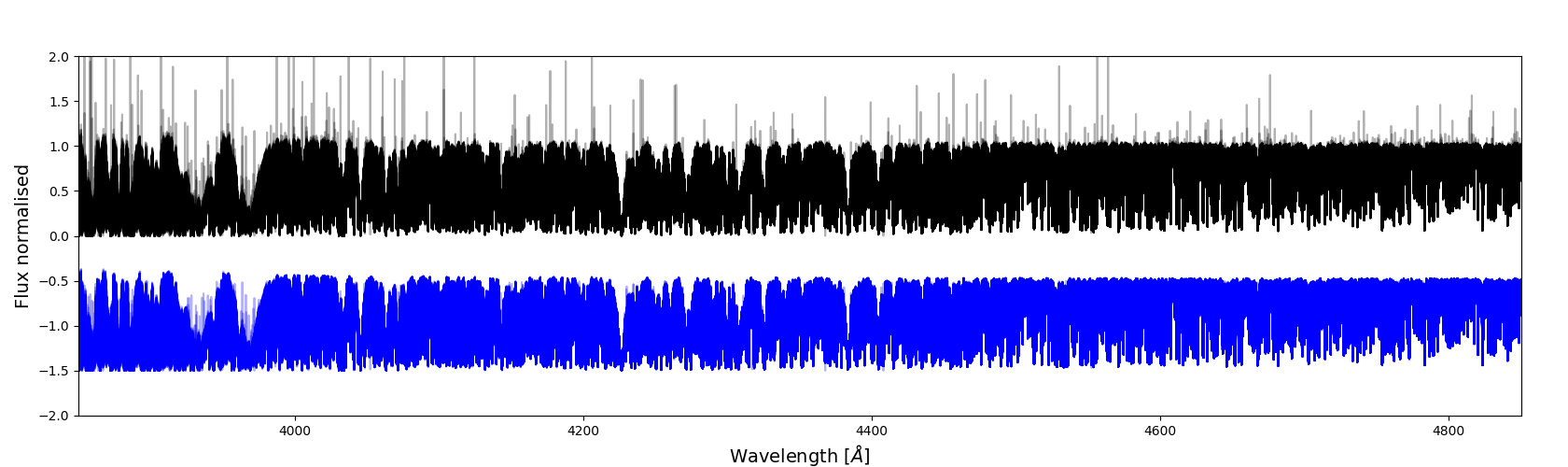}
	\caption{First flux correction performed by YARARA to remove cosmic rays contamination. The 277 RASSINE-normalised spectra of HD215152 are represented superposed before (\emph{black}) and after (\emph{blue}) YARARA cosmic rays cleaning. Cosmic rays are corrected by replacing every 5-sigma outlier larger than one by its value in the median spectrum.}
	\label{FigCosmics}
\end{figure*}

To be efficient, spectra time-series objects must be well continuum normalised, which is difficult due to the varying extinction of the Earth's atmosphere. A solution for echelle-grating instruments, consists in correcting the color by scaling the spectra with a reference template or reference color curve \citep[e.g.][]{Berdinas(2016),Malavolta(2017)}. However, such corrections are usually performed order by order which do not account for intra-order variations. A new approach based on alpha-shape regression was presented by \citet{Xu(2019)} on EXPRES spectra, where the authors demonstrated the better performance of alpha-shape algorithms compared to more naive approach as iterative fitting methods \citep{Tody(1986), Tody(1993)}. Similarly, we developed a Python code called RASSINE \citep{Cretignier(2020b)} and demonstrated its precision in flux on HARPS merged 1D spectra down to 0.10\%. RASSINE therefore appears as a well-suited tool to normalise spectra at the precision required to correct for the instrumental systematics. 
 
The initial products given as input to YARARA are normalised merged 1D spectra.  In this work, we started from HARPS merged 1D spectra produced by the official data reduction sofware (DRS), pre-processed them as follow, and then continuum normalised them using RASSINE. All the spectra are shifted in the stellar rest frame, night-drift corrected and evaluated on the same wavelength grid by a linear interpolation. Wavelengths longer than 6835 \ang{} were discarded due to the presence of the strong $\beta$ oxygen band. If the data contains large  Doppler shift variations, typically larger than $\sim$5 m/s on HARPS, induced by binary companion or massive planets, the spectra are shifted to cancel this RV shift before being interpolated on the common wavelength grid. Smaller RV variations do not need to be corrected for since smaller RV amplitudes do not produce strong flux signatures in river diagram. Such a correction was unnecessary for the three stellar systems studied in Sect.~\ref{sec:results}. Spectra are then nightly stacked to improve the S/N and finally normalised in automatic mode by RASSINE using the spectra time-series option, which allows to select the same local maxima on all the individual spectra (see \citet{Cretignier(2020b)} for more information).

We note that we decided to bin the spectra within a day, to help the normalisation process, as the upper envelope approach used by RASSINE to find the continuum is biased at low S/N. In addition, nightly binning helps in mitigating granulations and stellar oscillations \citep{Bouchy(2001b),Vasquez(2005), Nordlund(2009), Rieutord(2010b), Dumusque(2011e),Cegla(2019)}, which induce RV variations with time-scales smaller than a day. 

%If planets with periods shorter than 1-day are present, they should be added into the Keplerian model exactly as large RV amplitude planets are, but it was not the case here for the systems studied.

\subsection{Cosmic rays correction \label{sec:cosmics}}

Cosmic rays are high-energetic particles which produce straight bright lines on raw images.% and are flagged by the HARPS DRS using a Horn detection algorithm (\xav{source and check}). 
When the anomaly created by a cosmic falls on the same pixel as the stellar spectrum, the real stellar flux is lost and the extracted stellar spectrum is significantly affected. This will thus strongly affect the derivation of RV locally on the spectra, as we will see further\footnote{ We note that when deriving RVs using the cross correlation function technique \citep[CCF,][]{Baranne(1996),Pepe(2002)}, the perturbations induced by cosmic rays are generally averaged out.}.  In addition, in the context of YARARA, cosmic rays are seen as strong outliers and their impact has to be corrected for to improve the convergence of the different regressions that YARARA performs in the hereafter recipes. 

In \citet{Cretignier(2020b)}, the authors demonstrated that RASSINE is insensitive to cosmic rays, since they can be flagged as local maxima with excessive derivative. Thus, cosmic rays will appear as spikes with values higher than one in normalised flux units (see Fig.~\ref{FigCosmics}). To detect them, for each wavelength column, we performed a 3-sigma clipping on the flux of all the spectra\footnote{We note that rather than using a standard deviation to perform the sigma clipping, we used 1.48 times the median absolute deviation to be less biased by outliers.} and only removed the outliers larger than one. Such a technique removes efficiently the highest contaminations as visible in Fig.~\ref{FigCosmics}. We note that more outliers could be removed by pushing down the threshold on the sigma clipping, however, this could also remove other contaminations than cosmic rays, and removing them at this stage would make it more difficult to correct for them afterwards.
 %We did not remove every outliers regardless of their absolute flux value, since such approach could also remove spuriously other contaminations present in the barycentric-Earth RV (BERV) rest frame which reduces afterwards the efficiency of PCA algorithms that will be dedicated to those contaminations. Such criterion could also have the disadvantage to partially correct emission telluric lines in the infrared which is not the case here since HARPS spectra end at 6835 $\AA$. %To counter act this issue, it would be sufficient to adapt the criterion by imposing the outliers to be detected in both stellar and BERV rest frame, since an emission telluric line would a priori not be flagged in the second one. If the outliers exhibit a flux value smaller than 1, this latter can be considered as a small contaminant less likely to produce bad fit when multi-linear regressions will be performed. Remaining outliers will be cleared at the end of the pipeline} (see  Sect.~\ref{sec:thar}).

\subsection{Interference pattern correction \label{sec:fringing}}

\begin{figure*}[h]
	\includegraphics[width=18.5cm]{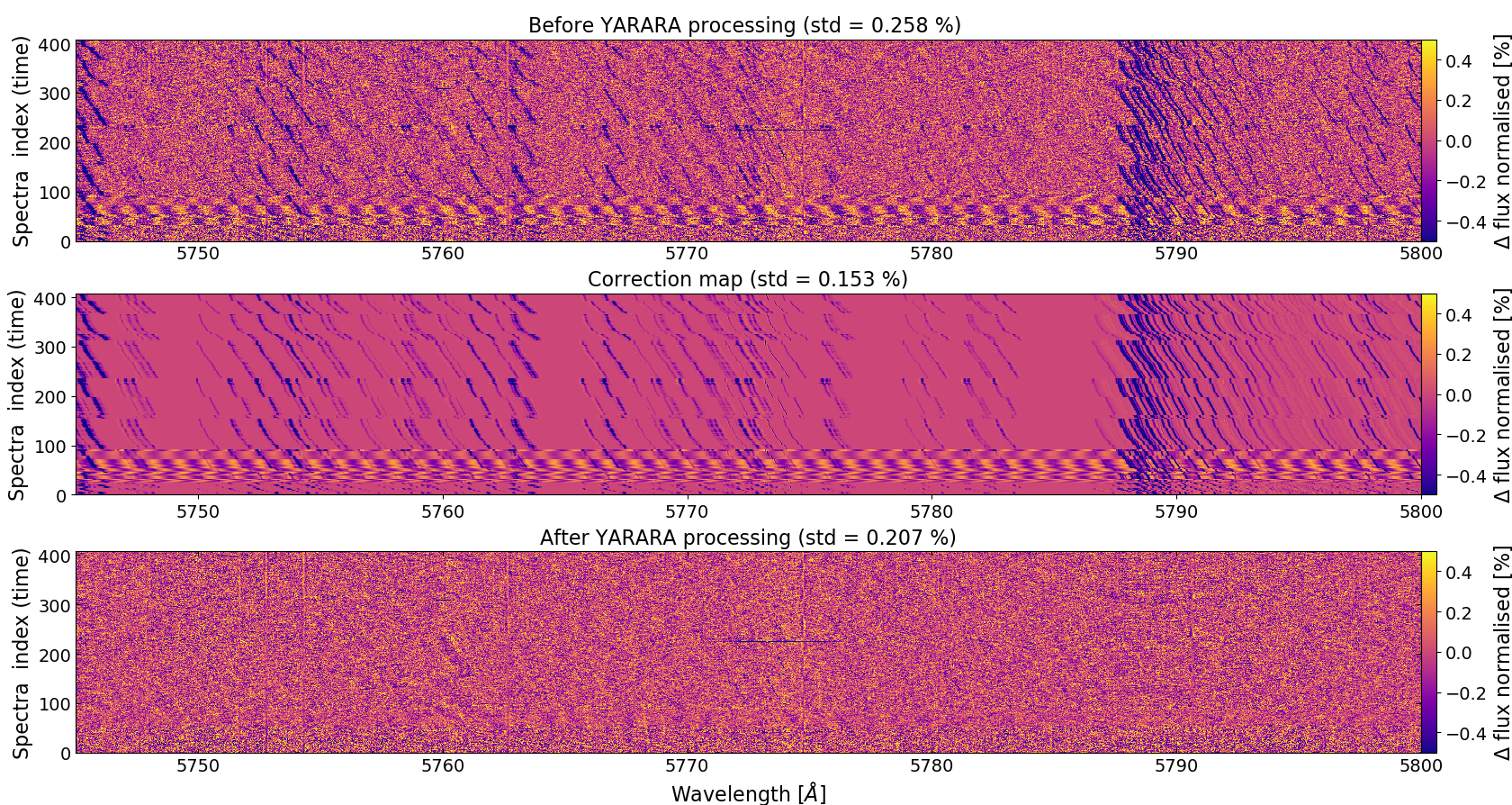}
	\caption{Second and third flux corrections performed by YARARA to remove interference patterns and telluric absorption lines contaminations on HD10700. The standard deviation of each residual map is indicated in the title. \textbf{Top:} River diagram before YARARA corrections. Spectra are suffering from interference patterns on two observational seasons (from index 30 to 97). \textbf{Middle:} Model fitted by YARARA for the contaminations. \textbf{Bottom:} Residual river diagram after YARARA correction.}
	\label{FigTelluric}
\end{figure*}

 On HARPS, an interference pattern was produced by mistake on FLAT calibration frames due to the use of a black body minus filter to balance the spectral energy distribution of a new Tungsten lamp installed in 2007. This filter was positioned in a collimated beam path, which produced the observed pattern following the thin-film interference law. This issue was solved during an upgrade of the instrument in August 2009, by moving the filter in a diverging beam. Hence, these systematics are only visible during two years of the HARPS observations (see Fig.~\ref{FigTelluric}). %As a comment, a similar pattern occurs on HARPS-N spectra but this latter have never been corrected so far from this effect. 
If the pattern frequency is low enough, RASSINE will naturally correct the oscillation as shown for ESPRESSO spectra in \citet{Allart(2020)}. If the pattern has a high frequency like for HARPS %or HARPS-N 
, with a periodicity shorter than one Angstr\"om, it needs to be corrected for.

The interference pattern has a unique frequency with time; however, the phase of the signal can change from day-to-day. We therefore decided to correct for the perturbing signal in the Fourier domain. To highlight at best the effect, we computed spectra ratio between each spectrum and the median of all excluding from the median all the spectra between $3^{rd}$ April 2007 and $24^{th}$ July 2009, which are known to be affected. We then computed the fast fourier transform (FFT) of all the spectra ratio. For the spectra ratio that are showing excess of power around a width\footnote{The interference periodicity can be found with the free spectral range law : width $= \frac{\lambda^2}{2n\Delta\lambda}$} of 1.5 mm for a refractive index $n=1$ \citep[see Fig.9 in][]{Cretignier(2020b)}, we replaced the region with the excess of power by the same region seen in the FFT of the highest S/N ratio spectra obtained outside of the contaminated time frame. The cleaned spectra are recovered by performing an inverse FFT followed by a multiplication with the median spectrum. This strategy successfully worked to remove the interference pattern as displayed in Fig.~\ref{FigTelluric}. We note that on HARPS, we needed to treat differently the blue and red detectors, as the gap in wavelength between the two CCDs was producing artefacts in the Fourier space.%We preferred to replace the power by those of the ratio spectrum free of the contamination rather than suppressing them since those frequencies can be relevant to model true variation in the ratio spectrum. In addition, this option was providing the best visual results and the least artefacts after the inverse FFT was taken. 

%The Fourier space therefore appears as the privileged place to highlight the contamination. We apply a correction separately for the blue and red detector since on HARPS a gap between the two CCD would produce artefacts in the Fourier space. The fringing is corrected by computing a ratio between all the spectra and the median spectra of the time-series, where the median is constructed by excluding all the spectra concerned by the fringing effect. This range being known and corresponds to all spectra between $3^{rd}$ April 2007 and $24^{th}$ July 2009. The ratio spectrum out of this temporal range with the highest S/N is kept as a reference model. A fast fourier transform (FFT) is applied on each ratio spectra in order to highlight the excess of power around 1.5 mm. The frequencies showing a power excess around this value are flagged and their power is replaced by the power of the reference model. Spectrum are recovered by performing the inverse FFT and after the product with the median spectrum is taken. This strategy successfully worked to remove the interference pattern as displayed in} Fig.~\ref{FigTelluric}.

\subsection{Telluric lines correction \label{sec:telluric}}

Telluric lines are terrestrial atmospheric lines fixed in the barycentric Earth RV (BERV) rest-frame which shift with respect to stellar lines and cross them due to the Earth reflex motion around the Sun, reducing the RV precision \citep{Artigau(2014),Cunha(2014)}. In the visible, telluric lines are exclusively coming from water vapour or oxygen lines and both species will be corrected for differently. Indeed, whereas the latters are stable with time and do not show strong depth variations, but only minor variations with airmass, water vapour lines exhibit large night-to-night and even intra-night variations \citep{Li(2018)} depending on the humidity content in the air and the atmospheric conditions. %As a comment, such short term variations are the main reason why a standard star close in time to the stellar observations has to be used when this method of correction is chosen. 
Some methods based on line profile telluric modelling \citep{Blake(2011),Bertaux(2014),Smette:2015aa} try to remove this contamination by using some prior information as those measured at the observational site \citep{Baker(2017)}. Unfortunately, such measurements are only providing local information on the ground, whereas the real information should be the column density of water along the direction of the observations, which is not a commonly derived quantity \citep{Kerber(2012)}. Another method consists in getting a high S/N spectrum of an hot standard star in order to extract the telluric spectrum \citep[e.g.][]{Artigau(2014)}, but such method is time consuming and observationally expensive. 

A more data-driven approach will take advantage that telluric lines are fixed in the BERV rest-frame and can therefore be disentangled from the stellar spectrum. This method has been successfully used in \citet{Bedell(2019)}, where the authors used a flexible
data-driven model to remove the contamination. We will use a similar approach, except that a PCA correction will solely be used as a second-stage correction, after applying a first step correction consisting in a multi-linear regression of the telluric CCF moments.

 We use as prior information the position of the water and oxygen telluric lines as given by Molecfit \citep{Smette:2015aa}. With these positions, a binary mask of water telluric lines is constructed in order to compute a telluric CCF. As a remark, the $\delta$ oxygen band around 5790 \ang{} is missing from the Molecfit database. We added the lines' positions by hand as they were clearly visible in our river diagram maps (see Fig.~\ref{FigTelluric})
 
 Because telluric lines present shallower depth than stellar lines, performing a CCF directly on the spectra will mainly contain stellar information. Switching to transmission spectra, by dividing the observed spectra by a telluric-free stellar spectrum $f_0(\lambda_j)$ allows to be less affected by stellar lines. 
 Since the telluric lines' density is quite low for the visible, we can build such a telluric-free stellar spectrum from the observations. This requires that for each specific wavelength column $\lambda_j$, at least one observation is not affected by tellurics. This hypothesis is equivalent to asking that the observations are probing a wide range of BERV values, where wide mean a range larger than the full width at half maximum (FWHM) of the telluric lines at the instrumental resolution. For HARPS, the FWHM is close to 3 km/s and therefore requires a minimum equivalent BERV span coverage.
 
 To perform our first level correction by CCF moments, a master stellar spectrum free of telluric lines is built by a masked median of the spectra-time-series, where a mask fixed in the BERV rest-frame is applied to hide the telluric lines according to our database. The binary mask of water telluric lines is used to compute the CCF of the spectra time-series divided by the master spectrum. %This master stellar spectrum is then used along with the binary mask of  water telluric lines to compute the CCF of the spectra time-series. 
 All the lines of the mask have the same weight, meaning that the obtained CCF can be seen as an average telluric line. Since its equivalent width (EW) can be interpreted as a proxy for water column density, this later is at first order proportional to the flux correction to apply. We note that a similar flux corrections is applied in \citet{Leet(2019)} with the SELENITE code, where the modelled correction is using as prior information both precipitable water vapor content and airmass.
 
 To correct for telluric contamination, we shifted the ratio spectra in the BERV rest frame and performed a multi-linear regression (see Eq.\ref{eq1}) by fitting both the contrast and FWHM time-series of the telluric CCFs on each wavelength column. The correction was performed only if the Pearson coefficient between the flux column and the contrast or the FWHM was higher than 0.4. We note that telluric CCFs can still be contaminated by stellar lines and their moments can therefore be corrupted. This can fortunately be detected as a mismatch between the telluric CCF RV and the BERV. When those two quantities were diverging by more than 1 km/s, we replaced the value of every moment by the median of the corresponding time-series. 

 After performing this first correction, residuals were visible in telluric line cores, in particular at times where the telluric CCFs were really deep, that can be interpreted as a non-linear behaviour of water lines when the humidity content in the atmosphere is too high. We then decided to apply a second correction based on a weighted PCA\footnote{The python code is freely accessible on GitHub : \url{https://github.com/jakevdp/wpca}} \citep{Bailey(2012)}. We performed a PCA on the previous residual matrix at the locations of clear telluric detection which is estimated to be telluric deeper than 0.10\% (see Appendix \ref{app:modeling2}). The PCA was weighted by the inverse of the flux uncertainty squared in order to prevent biasing the PCA algorithm with unusual low S/N observational epochs. The obtained PCA vectors were then fitted on each wavelength column of ratio spectra (see Eq.\ref{eq1}). Only the three first principal components were fitted, as adding more components was not reducing the variance further. The modeled telluric time-series was extracted by taking the difference between the raw ratio spectra time-series (in the terrestrial rest-frame) and the residuals after CCF moments and PCA components fitted. The telluric time-series model was then shifted back in the stellar rest-frame and use to correct the ratio time-series. The corrected spectra, in classical flux units, were recovered by multiplying the corrected ratio time-series by the telluric-free spectrum $f_0(\lambda_j)$. 
 
 The spectra time-series after this cleaning process can be seen in the lower panel of Fig.~\ref{FigTelluric} and we observe that the adopted methodology successfully remove telluric contamination at the precision level of around 0.21\%. The second stage of correction can be seen as a refinement of the telluric correction and will mainly correct for residuals let by the first stage correction. This can be appreciated on the individual components extraction performed on HD10700 (see Fig.~\ref{FigComponents} in Sect.~\ref{sec:hd10700}) for which the RV amplitude of the second stage correction is one order of magnitude lower. 
 
 The same approach was used to correct for oxygen lines. On a final note, the method described here will certainly fail in the infrared where the lines density of water is too high. A possible solution would be to perform a first order correction by telluric lines modelling \citep{Seifahrt(2010), Ulmer-Moll(2019)} before applying the PCA as a second-stage correction on the residuals. 

\subsection{Point spread function variation and activity corrections \label{sec:activity}}

The stellar CCF moments, such as the bisector inverse slope (BIS), the FWHM, the contrast or the equivalent-width (EW) have been widely used in the past to probe stellar activity \citep[e.g.][]{Queloz(2001),Povich(2001),Boisse(2011)}. In particular, since those moments are known to contain no Doppler information, they are still frequently used nowadays, to highlight power excess or deficit at specific periods in periodograms, when the purpose is to assess the nature of periodic signals. The CCF can be seen as a weighted average stellar line and its variation therefore represents the average behaviour of stellar lines. In the context of stellar activity, several works have shown that photospheric lines were changing in morphology due to stellar activity \citep{Stenflo(1977), Cavallini(1985), Basri(1989), Brandt(1990), Berdyugina(2003), Anderson(2010), Thompson(2017), Dumusque(2018), Wise(2018), Ning(2019), Lohner(2019)}. In \citet{Cretignier(2020a)}, we demonstrated how even a symmetrical flux variations, that should a priori not induce a RV variation, can still introduce a RV effect if the line profile is asymmetric, such in the case of a blended line \citep[see also][]{Reiners(2013)}. 

  \begin{figure*}[h]
	\includegraphics[width=18.5cm]{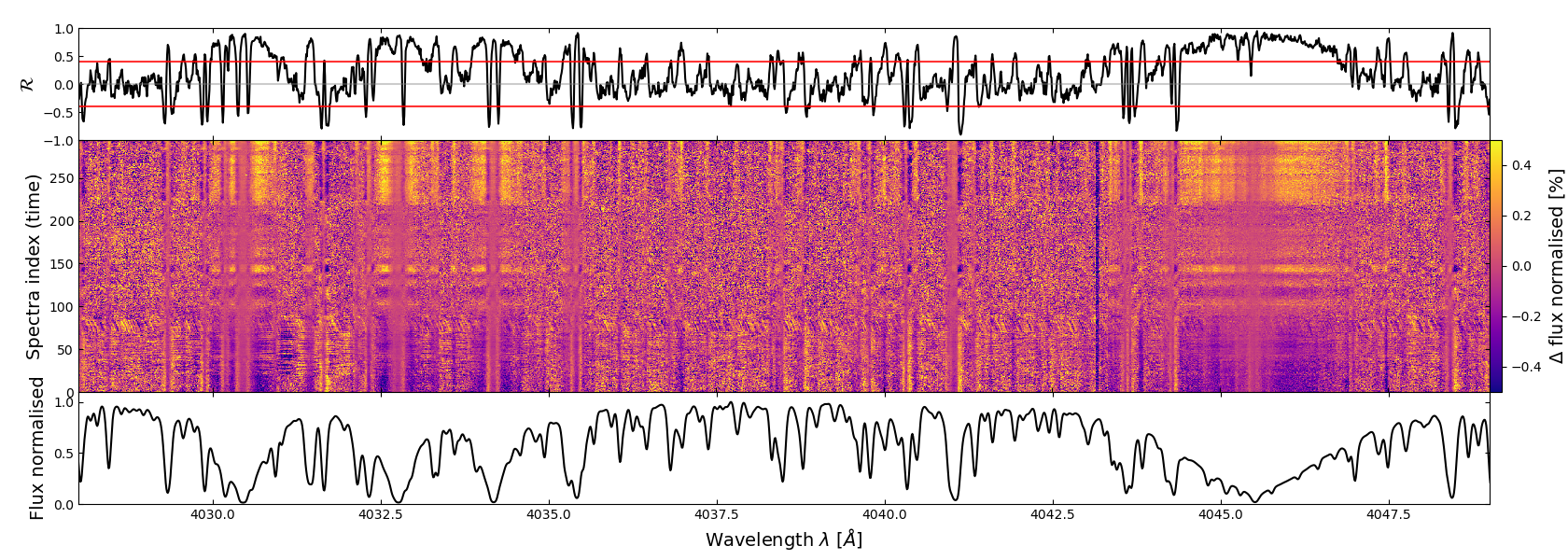}
	\caption{Illustration of the flux variations induced by stellar activity on $\alpha$ Cen B and its correlation with the S-index on a small wavelength range in the extreme blue. \textbf{Bottom:} Master normalised spectrum of $\alpha$ Cen B. \textbf{Middle:} The river diagram of the five years of alpha Cen B observations. Some lines clearly show long trend flux variations corresponding to the magnetic cycle of the star. \textbf{Top:} The Pearson correlation between the flux variations and the S-index show for several wavelengths a significant correlation with $|\mathcal{R}_{Pearson}|>0.4$ (red horizontal lines). Such correlations were already reported by previous studies for the 2010 dataset as \citet{Thompson(2017),Wise(2018),Ning(2019),Cretignier(2020a)}. The 2010 observational season is equivalent to index 102 to 150 on the figure for which the rotational modulation is clearly visible.%\textbf{Right:} Illustration that the correlation with the S-index is well explained by the stellar line profile. Line cores and continuum lay on the null gradient vertical line, whereas left and right wings present respectively negative and positive flux gradient values. Stellar activity is inducing simultaneously a core filling and line width broadening of stellar lines.}
	}
	\label{FigActivity}
\end{figure*}
 
  \begin{figure*}[h]
 	\includegraphics[width=18.5cm]{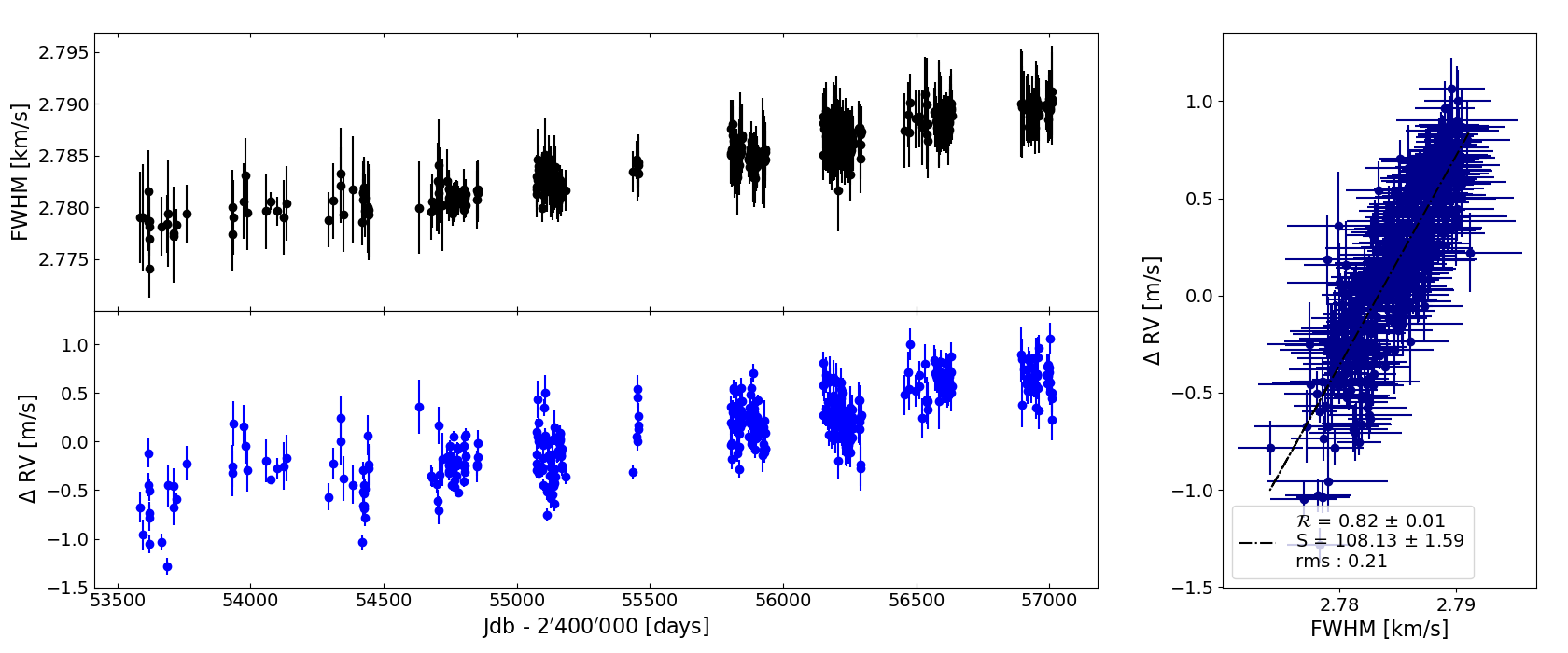}
 	\caption{Visualisation of the focus change due to HARPS ageing on the CCF FWHM and RV time-series of HD10700 obtained with the HARPS G2 mask. \textbf{Top left:} CCF FWHM time-series of HD10700. The star is known to be extremely quiet and the long-term trend observed is likely induced by the instrument ageing. The linear trend is about 1 m/s/year, which proves the impressive stability of HARPS (and HD10700) over a decade. A similar trend, however decreasing, is observed on the CCF contrast. \textbf{Bottom left:} RV difference obtained on the RV before and after applying the YARARA correction on the spectra-time-series. YARARA allows to suppress a long-term linear drift of 10 cm/s/year. \textbf{Right:} Correlation plot of the previous time-series, which shows the strong dependence between those two quantities, with a Pearson correlation coefficient of 0.82. %\st{The scaling factor between a change in FWHM and a RV change is about 108.1 m/km thus 0.1}. 
 	The 21 cm/s rms that we can see in the residuals can be explained by the simultaneous fit of the CCF contrast and CaII S-index along with the FWHM or bad fit on some wavelength columns.}
 	\label{FigFwhm}
 \end{figure*}

We note that in this paper, we will not use the information of the CCF moments to correct for stellar activity, as for the quiet stars that will be presented in Sect.~\ref{sec:results}, no excess power is visible in any activity proxy, and thus stellar activity is not an issue. CCF moments that are tracing line-shape variations unrelated to Doppler shift (namely contrast and FWHM) can be used to probe and correct the variations of the instrumental point spread function (PSF). To do so, we used an approach similar to the first-stage tellurics correction (see Sect. \ref{sec:telluric}), except that the spectra are analysed in the stellar rest frame and spectra differences are used instead of spectra ratio as for tellurics. The stellar CCFs are extracted from the G2 or K5 CCF mask of HARPS depending on the stellar spectral type and a multi-linear regression (see Eq.\ref{eq1}) of the contrast, the FWHM and the CaII S-index is fitted on each wavelength column.

\begin{figure*}[h]
	\includegraphics[width=18.5cm]{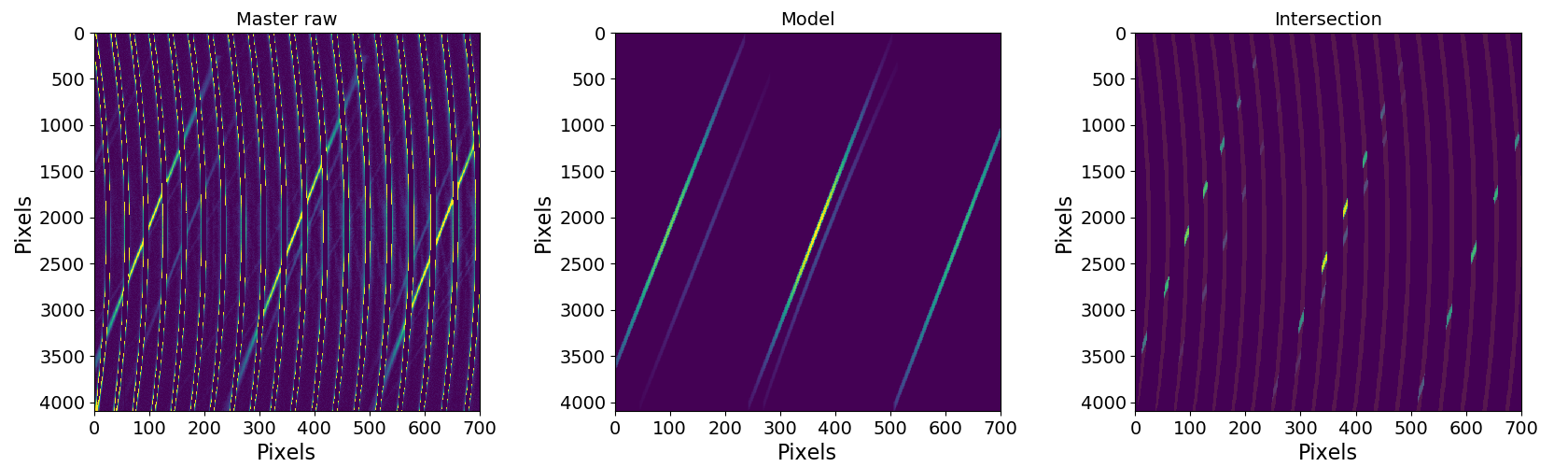}
	\caption{Derivation of the static product used in the fourth flux correction performed by YARARA dedicated to ghosts contamination. \textbf{Left:} A master raw image of the blue detector obtained after stacking two months of FLAT calibration frames when only the science fiber was illuminated. The echelle spectrograph orders are the vertical parabola which are masked here to highlight the ghosts visible in the background. \textbf{Middle:} Ghosts as fitted by our models. \textbf{Right:} Map of ghost contamination indicating the location of the ghost of the science fiber on itself. Bright color highlights the intersection of the ghosts with the physical orders. A similar map was derived for the ghosts of the simultaneous calibration fiber crossing the order of the science fiber.}
	\label{FigGhost}
\end{figure*}

The advantage to fit simultaneously the S-index and moments of the CCFs is that whereas the former should only contains stellar activity effects, the latters can contains simultaneously instrumental and stellar activity components. Fitting all of them therefore lead to lift the potential degeneracies that may exist. We can wonder if the fit in flux of the S-index is motivated. Despite the higher sensitivity of the CaII\,H\&K lines to active regions, it is known that all photospheric lines change in depth with stellar activity \citep{Basri(1989)}. Even if photospheric and chromospheric variation can differ due to their different formation layers, several deep broad lines are also formed in the chromosphere as the CaII lines and will therefore undergoes similar variation with different strengths. This is confirmed in Fig.~\ref{FigActivity} where we represented the five years of observations of $\alpha$ Cen B and the correlations of each wavelength column with the S-index. Several wavelength columns show a significant correlation ($\mathcal{R}>0.4$) with the S-index. %By expressing the spectrum in the line profile space, obtained by switching from the wavelength $\lambda$ variable to the gradient of the flux variable $df/d\lambda$, we observe that the line cores show positive correlation and lines wings negative correlations. 
The modifications observed are equivalent to line core filling and line width broadening which are effects already known from stellar activity on the 2010 dataset \citep{Thompson(2017),Wise(2018),Ning(2019), Cretignier(2020a)}. We do not investigate further stellar activity since such an analysis is left for a future dedicated paper and the targets selected hereafter are quiet stars.

It is important to mention that CaII\,H\&K lines can be contaminated by ghosts (see Sect.~\ref{sec:ghost}) which are not yet corrected for at that moment. Hopefully for HARPS, as demonstrated in Appendix \ref{app:modeling}, ghosts never fall simultaneously on the CaII\,H and CaII\,K lines. In addition, the contamination depends on the systemic RV ($RV_{sys}$) of the observed star that will shift the spectra and thus the position of the cores of the CaII\,H\&K lines relative to ghosts that are fixed in the barycentric rest-frame. Since both chromospheric lines react in a similar way, the activity proxy that we fit will be either made of one line (CaIIH or CaIIK) or both combined (S-index) depending on the $RV_{sys}$ of each star (see Fig.~\ref{FigContamGhost}). For an instrument strongly contaminated by ghosts, a careful extraction of the S-index should be performed \citep{Dumusque(2020)}. 

%\st{This recipe was motivated by the following observation. It is interesting to note that no matter how much the star is active or how much the star is inactive, the stellar CCF moments (other than the first one) are tracing line shape variations not related to a Doppler shift. As a consequence, the contrast, FWHM or EW of the stellar CCF can be used to correct the flux variations which induce a change in the line profile. It is similar to the first order telluric lines correction, excepted that this correction is now performed in the stellar rest frame.}

\begin{figure*}[h]
	\includegraphics[width=18.5cm]{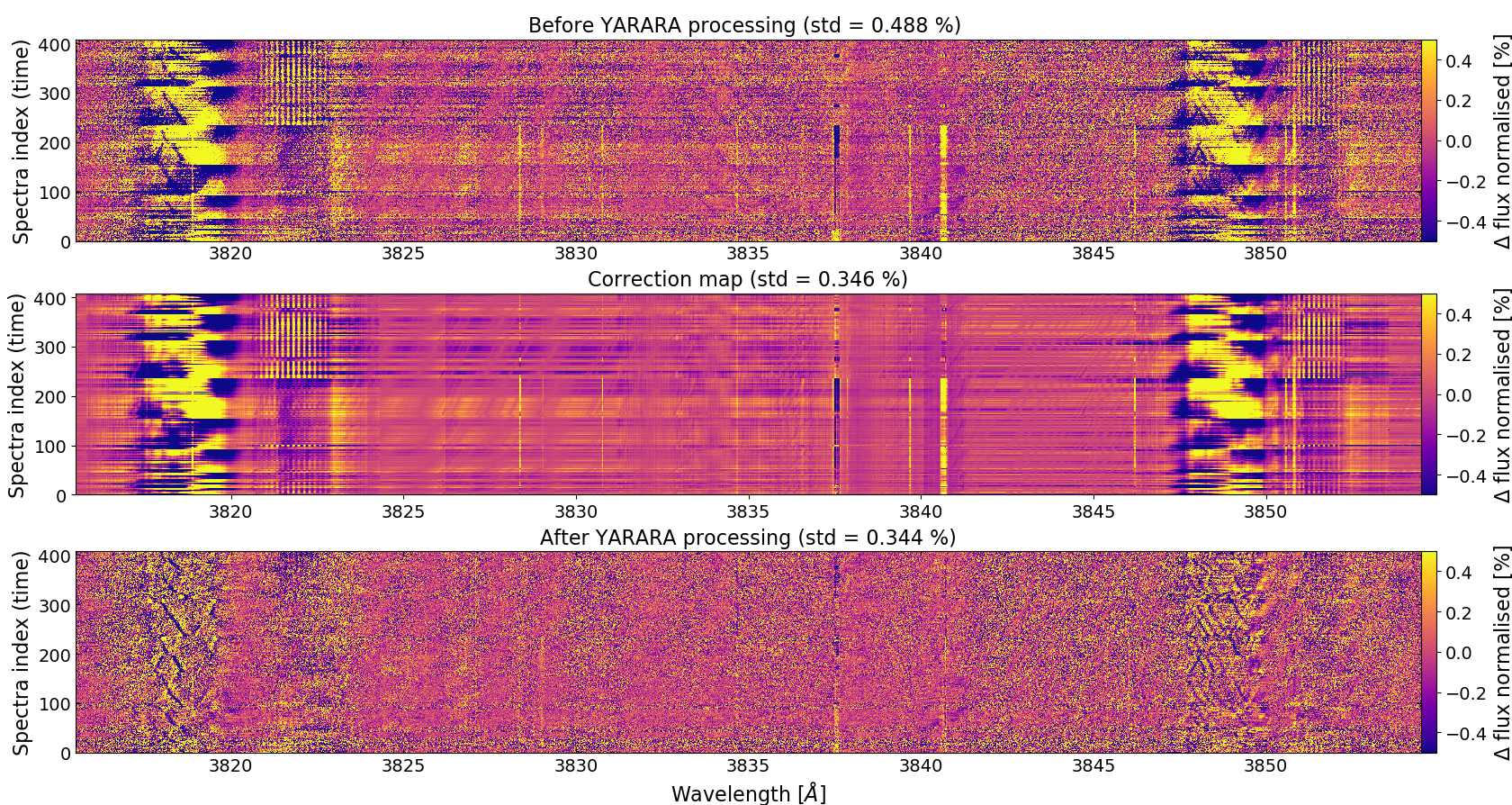}
	\caption{Ghost, Thorium-Argon and continuum corrections performed by YARARA on HD10700. The river diagram is produced in the BERV rest frame. \textbf{Top:} Spectra-time-series before YARARA corrections. Spectra are suffering from Thorium-Argon contamination (bright vertical lines, for example close to 3840 \ang{}), and Fabry-Pérot contamination by a ghost of the simultaneous calibration fiber (vertical bright comb around 3822 and 3851 \ang{}). The strong contamination around 3819 and 3849 \ang{} are due to two ghosts of the science fiber. \textbf{Middle:} Model of the contamination fitted by YARARA.  \textbf{Bottom:} Residual river diagram after YARARA correction.}
	\label{FigContam}
\end{figure*}

When analysing the data of the extremely quiet star \object{HD10700}, it is possible to observe the change of HARPS' focus as a function of time, which is associated to the ageing of the instrument. Indeed, the CCF FWHM (top left panel of Fig.~\ref{FigFwhm}) and the contrast show a net increase and decrease with time, respectively. This is coherent with a slight decrease of HARPS resolution with time, equivalent to a PSF broadening. The EW, which is at first order the product of the FWHM and the contrast, is not showing such a long-term trend. It is however relevant to note that the FWHM variation observed is only about 1 m/s per year. It confirms the remarkable stability of the instrument along its lifetime, as well as the incredible stability of the star itself.

To measure the RV impact of the instrument ageing, we computed the RV before and after applying the flux correction using the HARPS G2 cross-correlation mask and measured the difference between both RV time-series. This latter is displayed in the bottom left panel of Fig.~\ref{FigFwhm} and is strongly correlated with the FWHM variation as we can see in the right panel of the same figure, with a Pearson correlation coefficient of 0.81. The RV impact of the instrument ageing is extremely small, with a linear slope of only 10 cm/s/year. We note that even if such a linear correlation between FWHM and RV can theoretically be used to correct for instrumental ageing on the RV time-series, correcting for the effect at the flux level is a more robust way of handling this issue.
%\st{We note that the residuals rms between the FWHM and the RV correction is of 21 cm/s. Those residuals can be explained by the fact that along with the FWHM, we also simultaneously fit the CCF contrast and the CaII S-index on the wavelength columns and because in some cases the multi-linear fit is not well converging due to outliers.
%Even if such a calibration value can theoretically be used to correct instrumental ageing (which is already done in the actual HARPS DRS), we point out that such a strategy of spectra time-series flux correction is far more motivated than the correction of the RVs values. }

\subsection{Ghosts correction \label{sec:ghost}}

Ghosts are spurious reflections of the physical orders occurring inside the instrument which produce tilted-duplicated images of the spectral orders on the CCD. Due to their reflection nature, ghosts are at maximum a hundred times less luminous than the original echelle orders, as measured on a master FLAT raw frame (see Fig.~\ref{FigGhost}). As a consequence, ghosts are difficult to distinguish from the read-out-noise of the detector on individual calibration frames, and no calibration frames can be produced to highlight "only" ghosts since those are secondary reflection of the main echelle orders. However, ghosts appear always at the same position on the detector and can thus be highlighted on high S/N images obtained after stacking several raw frames together. As an example, we show on the left panel of Fig.~\ref{FigGhost} an image obtained after stacking two months of HARPS FLAT calibrations raw frames and we clearly see the contamination induced by ghosts as oblique bright lines.

The ghosts' contamination is not corrected in the HARPS DRS because it is difficult to deal with it at the extraction level due to the relative small angle between ghosts and the main echelle orders, but mainly because when observing a star, the ghosts will be a copy of the stellar spectrum itself, and not the "flat" spectrum of a tungsten lamp used for flat-fielding. To our knowledge, no study on HARPS have analysed to which main echelle orders the ghosts were belonging to, thus making any correction extremely difficult. 

Even though difficult, a proper way of correcting for ghost contamination in the context of RV measurements was never investigated further, as ghosts only contaminate spectral regions with very low flux, like the extreme blue in stellar observation, and those regions do not contribute significantly to the extracted RVs.
However, ghosts can significantly contaminate the cores of the CaII H\&K lines in the extreme blue, at 3834 and 3969 \ang{}, which are used to derive the calcium activity S-index, the main stellar activity indicator for G-K dwarfs \citep{Wilson(1957),Oranje(1983b),Baliunas(1988)}. This is the case in the HARPS-N solar observation, where a ghost correction technique was performed to derive a calcium activity index free of contamination \citep[see][]{Dumusque(2020)}.

To be able to derive a calcium activity index free of systematics, but also to obtain meaningful line-by-line RV information in the extreme blue, we decided to correct for ghost contamination at the flux level with YARARA. The first step consists in modelling the ghosts on the raw frames (middle panel) in order to determine their intersection with the science orders (right panel). The modelling of ghosts is detailed in Appendix \ref{app:modeling}. To obtain the wavelengths of a spectrum that are contaminated by ghosts, we took advantage of the HARPS instrumental stability, which allows us to use a static wavelength solution to map contamination from pixels to wavelengths. As the contamination is additive, we looked for the contamination in spectrum difference time-series, obtained by subtracting to each spectrum a master built as the median of all spectra. The river diagram map of spectrum differences shifted in the BERV rest frame is displayed in the top panel of Fig.~\ref{FigContam}. Around 3819 and 3849 \ang{}, we see ghost contamination from the science fiber (Fiber A), and around 3822 and 3851 \ang{}, ghost contamination from the reference fiber (fiber B). We then selected, at the location of ghost crossings, all the wavelength that where showing a relative flux variation more important that 1\% and trained a weighted PCA \citep{Bailey(2012)} to model the ghost contamination. Finally, we fitted for all the wavelengths columns located in a ghost crossing region, the first three PCA components for ghosts of fiber A and the first two for ghosts of fiber B (see Eq.\ref{eq1}).

We note that around 3819 and 3849 \ang{}, the ghost contamination of fiber A present an emission/absorption pattern which is unexpected. Naively, we would expect only an emission like for Thorium-Argon contamination, however on a much larger region (see top panel of Fig.~\ref{FigContam} and Sect.~\ref{sec:thar}). As this is not the case, one explanation is that the master spectrum subtracted is not free of ghosts. Indeed, since ghosts cross the orders with a small penetration angle ($\sim$5$^\circ$), their contamination is present over about one hundred pixels. In consequence, even if the star studied present large BERV values (at maximum $\sim$60 km/s, thus 73 HARPS pixels, as a pixel is $\sim$820 m/s), significant wavelength ranges are always contaminated by ghosts. An option would be to optimize the way the reference master is built, however, such a solution will not work for stars with a small span in BERV. It is important to note that this issue will provide an inaccurate correction, however, still very precise, which is the important factor when deriving RVs. %\st{Therefore, we decided to simply add one extra component of the PCA}. 
The same problem does not occur for ghost contamination of fiber B since this fiber is illuminated either by a Thorium-Argon lamp or a Fabry-P\'erot etalon that present distinct emission peaks, and not a continuous emission like the ghosts of fiber A, which are reflections of a stellar spectrum. This explains why we fitted the first three PCA components for fiber A, and only the first two for fiber B

\begin{figure*}[h]
	\includegraphics[width=18.5cm]{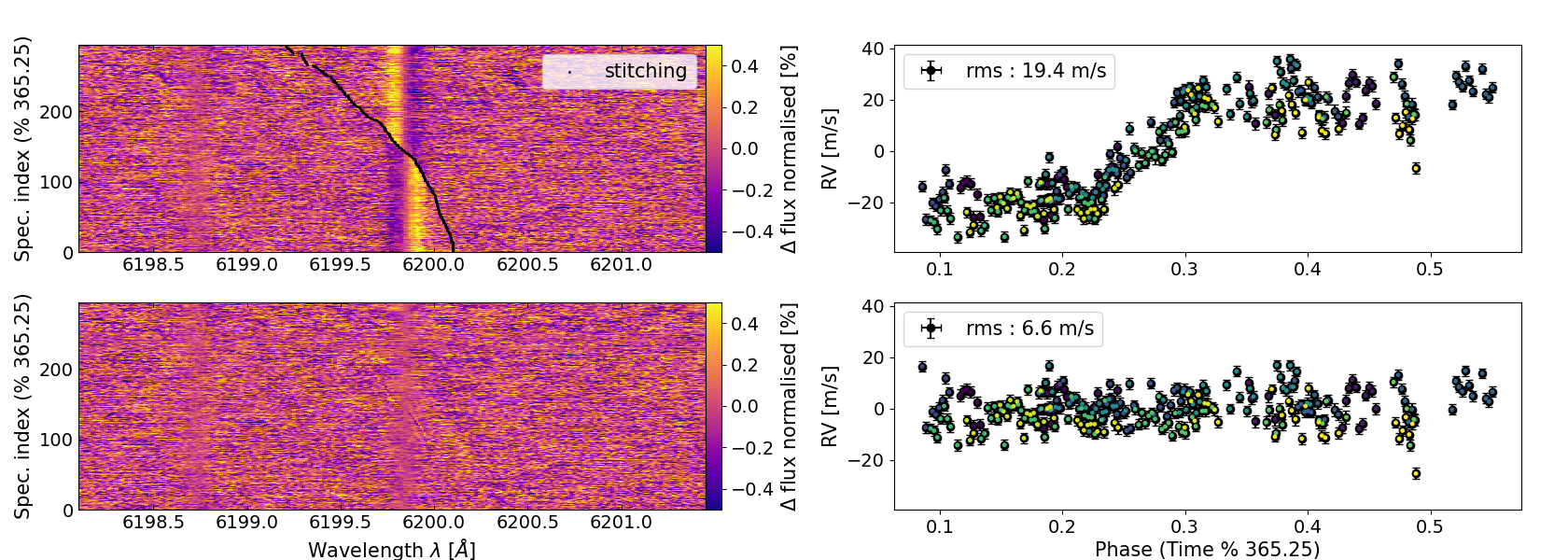}
	\caption{Stitching correction performed by YARARA on the five years of HD128621 data for the 6200 \ang{} stellar line described in \citet{Dumusque(2015)}. The time dimension is phase-folded at 1-year. \textbf{Top left:} Spectra-time-series before YARARA corrections. The position of the stitching is indicated by the black curve and is used to mark the delimitation of the step function used for correction (see text). \textbf{Bottom left:} Same as top left after YARARA processing. \textbf{Top right:} LBL RV of the line crossing the stitching. Color encodes the unfolded time.  \textbf{Bottom right:} Same as top right after YARARA processing. The RV rms has been reduced from 19.4 m/s down to 6.6 m/s. LBL morphological correction (see Sect.~\ref{sec:lbl}) mitigates the rms even further, down to 5.6 m/s. }
	\label{FigStitching}
\end{figure*}

%\st{The same strategy is adopted for ghosts of fiber A and B, with the only exception that for fiber A, we fit for the first three components, while only the two first ones for fiber B. We perform the analysis in the BERV rest frame and train the PCA on wavelengths contaminated by a relative intensity higher than 1\%. We kept the three first components of the fit for ghosts A, against two components for ghosts B. Then, a multi-linear regression is performed with the vectors on all the ghost regions flagged by the static calibration product. }

The principal component model of ghost contamination from fiber B clearly captures the Fabry-Pérot structure as visible by the emission comb around 3822 and 3851 \ang{} in the middle panel of Fig.~\ref{FigContam}. This structure is only visible once the Fabry-Pérot has been routinely used on HARPS, which happened in 2012. We can see few remaining residuals in the lower panel of the same figure inside the ghost A regions which appear as dark lines perfectly phase-folded at one sidereal year. Those lines are the stellar lines of the ghosts contaminating the science fiber. Since both the ghost stellar lines and the science stellar lines are shifted with the BERV, those lines are not fixed in the BERV rest frame and determining their rest-frame is out of the scope of the present paper. %, but rather in a rest frame equivalent to roughly 2 times the BERV which indicates that this ghost family is flipped compared to the main order. 

Although ghost contamination is strongly mitigated after YARARA correction, the structured residuals could still impact the RV of spectral lines appearing in the contaminated regions. We will see that those residuals will impact the derived RV of spectral line in a characteristic way different from a Doppler shift, and we will correct for those at a later stage (see Sect.~\ref{sec:lbl}).

%To understand it, let us assume that the ghost of a spectral order is perfectly parallel to the main order and is falling on the same physical order but its direction is flipped since it is a reflection. In the river diagram representation and in stellar rest frame, the ghost line would be visible as an absorption feature drifting with twice the BERV value. All ghost stellar lines are therefore fixed in a rest frame defined as a scaling factor times the BERV. However, the exact value of the factor depends, on the penetration angle of the ghost with the order as well as the wavelength of the order and of the ghost stellar line. Even if those information were in our possession, the correction would be quite tricky to accomplish since we do not have a model intensity for the ghost line. 
%\st{We therefore decided to solve this issue differently, since those residual variations can be disentangled from real Doppler shift in the LBL analysis} (Sect.~\ref{sec:lbl}). 

\subsection{Stitching correction\label{sec:stitching}}

Stitching are detector systematics induced when manufacturing a CCD. The HARPS 4096 $\times$ 4096 CCD is a mosaic composed of 8 blocks of 512 $\times$ 1024 pixels, which are not perfectly aligned together. The pixels located at the stitching of the different blocks have a different size than the pixels within blocks, due to manufacturing limitation. If not properly accounted for when extracting spectra and measuring a wavelength solution, this difference in size will induce a RV variation at a period of one year, as first shown by \citet{Dumusque(2015)}. In this paper, the authors strongly mitigate the induced one-year signal by simply rejecting from the RV computation the part of the spectrum affected by the stitching effect. However, this rejects a significant amount of RV information, and a better way to handle this problem is to measure the size of the pixels at the different stitching location and include it in the wavelength solution derivation \citep{Coffinet(2019)}. We will see below that stitching effect can also be efficiently corrected for at the flux level.

%\st{by the technical limitation during the fabrication of the HARPS CCD detector. The detector is composed of small chunks of 8 blocks of 512 pixels forming a mosaic of 4096 x 4096 pixels. In an ideal world, all the chunks would perfectly fit together, which was not possible due to technical limitations, producing small gaps between the chunks that are called stitchings. The stitching effect was first understood in} \citet{Dumusque(2015)}, \st{where the authors demonstrated that by excluding stellar lines crossing the stitchings, the 1-year signal visible in the periodogram of the HARPS data of certain stars was highly reduced. Since the wavelength solution implicitly assumes that pixels are equidistant, the gap is producing a wrong wavelength solution which is disturbing the stellar lines profile with a 1-year periodicity due to the BERV. Several ways exist to solve this issue, either by rejecting contaminated stellar lines} \citep{Dumusque(2015)} or by correcting the wavelength solution \citep{Coffinet(2019)}, \st{however we will see that stitchings are easily corrected with spectra time-series, since the model of the flux variation can be described by a step function.}

In Fig.~\ref{FigStitching}, we display the same stitching-crossing stellar line at 6200 \ang{} for HD128621 as reported in \citet{Dumusque(2015)}. The left panel show the flux anomaly created by the wrong wavelength solution at the stitching position before and after YARARA correction. We note that the time dimension in those river diagram maps is phase-folded at one year. The right panel of the same figure show the LBL RV of the 6200 \ang{} spectral line, phase-folded at 1-year, before and after YARARA correction. It is clear that before correction the RV of the line shows a sigmoid-like shape variation, with a jump of 40 m/s between phases where the stitching is on both sides of the spectral line. From the left panel of Fig.~\ref{FigStitching}, we can see that for each wavelength columns, the effect in flux is a step function, where the position of the step correspond to the stitching location. Instead of fitting for a step function which could be biased by outliers, we computed the median below and above the stitching boundary for each wavelength column crossing a stitching and subtracted this model. We clearly see that the stitching flux anomaly is strongly mitigated in the river diagram map residuals. When deriving the LBL RV of the 6200 \ang{} spectral line using the spectra corrected from YARARA, the original 19.4 m/s RV rms of the line is strongly reduced down to 6.6 m/s. 
%The main part of this remaining jitter is related to the stellar activity of HD128621 and will be described in a forthcoming paper. 
Like in the case of ghost contamination, the remaining small residuals can be corrected further when studying the LBL RVs as the induced effect will be different from a Doppler shift. Performing LBL morphological corrections (see Sect.~\ref{sec:lbl}) manages to correct the systematic down to a rms of 5.6 m/s.

\subsection{Thorium-Argon correction\label{sec:thar}}

A Thorium-Argon lamp was used on HARPS \citep[before the use of the Fabry-P\'erot etalon][]{Wildi(2011)}) as a reference calibrator to measure the drift of the instrument throughout the night. To do so, the light from the Thorium-Argon lamp was injected in the reference fiber (fiber B) of the spectrograph, to simultaneously measure the instrumental drift along with stellar observations. During a stellar observation, the reference Thorium-Argon spectrum on fiber B presents however a significant number of saturated lines, that "bleed" on the detector and contaminate the nearby science fiber. Lowering the flux of the lamp to avoid saturation is not an option, unfortunately, as otherwise the S/N of the Thorium spectrum would not be sufficient to reach the precision required to measure the tiny drifts of the spectrograph overnight.

The HARPS DRS corrects for this effect by using Thorium-Argon contamination calibrations. Those calibrations have only fiber B illuminated by the Thorium-Argon lamp. The contamination of the lamp at the position of the science fiber is extracted and then fitted for in stellar observations. Unfortunately, these calibration frames were only rarely produced on HARPS, which implies an imperfect correction, as lines intensities change with time due to ageing of the lamp, or due to the change of lamp which happened a couple of times throughout HARPS lifetime\footnote{an history of lamp change for HARPS can be found here: \url{http://www.eso.org/sci/facilities/lasilla/instruments/harps/inst/monitoring/thar_history.html}}. We can clearly see at wavelengths 3837 \ang{} and 3841 \ang{} in Fig.~\ref{FigGhost} that the use of static contamination products do not provide an excellent correction.

To perform a better correction, we first used the available Thorium-Argon contamination calibrations to select the spectral regions contaminated by saturated lines. We used a 1.5 interquartile (IQ) sigma-clipping to detect anomalous flux intensities compared to the noise level. The IQ being more robust statistically to detect outliers \citep{Upton(1996)}. We then trained a weighted PCA on the selected wavelength columns in the BERV rest frame, and corrected for the contamination by fitting the first two PCA components on all the wavelength columns. We decided to fit the model on the full wavelength domain instead of flagged regions like for ghosts, since detecting the contamination in the calibrations frames is difficult. To prevent the first two PCA components to fit for any type of systematics, we performed the correction only if the Pearson coefficient with one of both PCA vectors was higher than 0.4.

%\st{in order to define a mask at the wavelength contaminated by fiber B outflows. The said wavelengths were flagged by a sigma clipping which highlights anomalous flux intensities compared to the noise level. In the pipeline, a weighted PCA is trained on those wavelength columns in BERV rest frame, and the two first principals components are fitted on the full wavelength range of the spectra time-series. We decided to fit on the full domain instead of flagged region, in contrast for ghosts, since clearly the contamination product has a limited S/N which rends difficult the detection of all the contaminations. We therefore used the same strategy than for the water telluric lines correction, namely by accepting the correction only if the Pearson coefficient with one of both PCA vectors was higher than 0.4.}

\subsection{Continuum improvement and iterative processing \label{sec:smooth}}

So far, a recipe was always building its own master spectrum, free as possible of the contamination it was correcting for. Since spectra at the end of the pipeline are cleared at best of all the observed systematics, the final master spectrum is necessarily better that the one available at the beginning of the pipeline. For that reason, a second iteration trough the different recipes by using always the master spectrum obtained at the end of the first iteration is performed. As an example, the first correction that benefits from the new master spectrum is the Savitchy-Golay continuum fit performed in RASSINE. Indeed, since telluric lines are initially not corrected for, they induce a poor continuum normalisation around telluric bands.%\st{, which translates as wiggles instability}. 

After the second iteration, no significant flux variation is remaining in the river diagram maps, except small residuals at ghost crossing and a few outliers. To get rid of these remaining systematics, we finally perform a 2-sigma clipping on each wavelength columns and replace outliers by the median value of the corresponding wavelength column, as done for cosmic correction (see Sect.~\ref{sec:cosmics}), with the exception that no condition on the flux level is imposed.

%\st{We tested performing a third iteration, however, the improvement was negligible.}

%\st{It therefore makes sense to perform a second iteration through the different pipeline recipes, after a first iteration is finished. We tested that after finishing a first iteration of the pipeline, a second iteration . We only performed a second iteration, since third iteration was showing negligible improvement. The strategy therefore consists in reprocessing the input uncorrected spectra time-series with the master spectrum fixed in each recipe during the second iteration.} %As a comment, we could have choose to begin from the corrected spectra-time-series, but this choice would be rather unmotivated and bad fitting on wavelength columns would propagate to produce larger systematics which justifies our choice to begin again from the input, namely the uncorrected spectra-time-series.

 %\st{The full chain of recipes described previously is launched a second time. After that, no significant flux effect was remaining in the river diagram maps, excepted ghost lines and a few outliers. We therefore concluded the flux correction of YARARA by applying the same sigma-clipping as performed for cosmics correction , excepted that a 2-sigma clipping is applied and no condition on the flux level is imposed this time.}

\subsection{Stellar mask optimisation and LBL RVs \label{sec:mask}}

YARARA was developed to extract locally the RV information in a stellar spectrum, as done in %}\st{to be fully embodied in the LBL framework}
\citet{Dumusque(2018)}. One of the final purposes of the pipeline is to produce better LBL RVs, since spectra are precisely normalised thanks to RASSINE and spectra are cleared out of their main contaminations. A great advantage after YARARA post-processing is that the available master spectrum is expected to be almost completely free of all known contaminations. Such a master can thus be used to perform cross-correlation and derive a RV value per spectrum \citep{Anglada(2012),Gao(2016),Zechmeister(2020)}, or derive LBL RVs.

LBL RV extraction by first order flux linearisation \citep{Bouchy(2001)} only requires three quantities: 1) a master template used to measure the flux variation $\delta f$ between a spectrum and the master, 2) the line profile gradient $\partial f/\partial \lambda$ to account properly for the weighting of the RV information and 3) the window centered on a spectral line to extract its RV signals. A few surprising results can be found if a generic mask is used instead of a tailored lines selection in LBL RVs analysis. Indeed, due to blended lines which are in some way unique for each star depending on the stellar temperature, metallicity, $v \sin(i)$ or instrumental resolution, a generic mask will in some case %\st{completely miss to correctly extract the RV of the lines which will} 
produce spurious RV signals, which was demonstrated in \citep{Cretignier(2020a)}. We note that similarly to \citet{Dumusque(2018)} and \citet{Cretignier(2020a)}, \citet{Lafarga(2020)} showed that tailored lines selections, produced by their public code RACCOON, can be useful to boost the S/N and decrease the photon noise, which produces in general better RV precision.

We implemented in YARARA a similar line window selection as described in \citet{Cretignier(2020a)}. This consists simply in selecting for each line, the wavelength range between the two local maxima surrounding the local minimum formed by the line core defined as the local minimum. Doing so, blended lines will sometimes be grouped together but this is not an issue as soon as signals specific to each line profile do not try to be interpreted \citet{Cretignier(2020a)}. We note that the obtained selection of lines is not excluding spectral regions strongly contaminated by telluric lines, as commonly performed by generic line masks, and therefore the line selection of YARARA can be used to extract RV information with the highest S/N.

%\st{The master spectra is constructed by stacking the corrected normalised merged 1D spectra, weighted by their initial bolometric flux.In YARARA, the window for the LBL extraction are simply defined as the wavelength range between both local maxima surrounding the local minima formed by the line core. The result from this process is that each star possess now its own optimised stellar mask somewhat similar to the Fig.2 in} \citep{Cretignier(2020a)}. \st{Note that our stellar mask is not excluding lines contaminated by telluric lines, as commonly performed by generic mask. All the results displayed in} Sect.~\ref{sec:results}, \st{are therefore obtained without excluding any spectral region. The stellar masks constructed by YARARA are therefore the masks optimising the most the photon noise by construction.} %However, here we assume that all the contaminations will be correctly removed which is perharps not true for low S/N spectra, for this reason YARARA also derives in parallel the CCFs with the usual and generic HARPS mask closest to the stellar spectral type. 

If we want to use the obtained line selection as a mask for performing cross-correlation, we need to obtain the proper depth of the line relative to the continuum, as this is used as weight when deriving the CCF \citep{Pepe(2002)}. In addition, if we want to derive LBL RVs using this line selection, we also need to estimate at best the continuum and depth of the spectral lines, as the LBL technique uses the line profile gradient $\partial f/\partial \lambda$ as weight. In regions where a continuum exists, the normalisation by RASSINE allows to derive the proper estimate for the continuum and line depth. However, this is no longer the case in the blue part of solar-like spectra or in the spectra of M-dwarfs, as no continuum exists due to the strong absorption. This problem is solved by measuring the deviation of the estimated RASSINE continuum with respect to a stellar template close to the effective temperature of the star, and correcting for it, as shown in \citep{Cretignier(2020b)}. As an example, we shown in Appendix \ref{app:continuum} and Fig.~\ref{FigContiGJ654}, the YARARA correction for continuum opacity in the stellar spectrum of a M1.5 dwarf.

\begin{figure*}[h]
	\includegraphics[width=18.5cm]{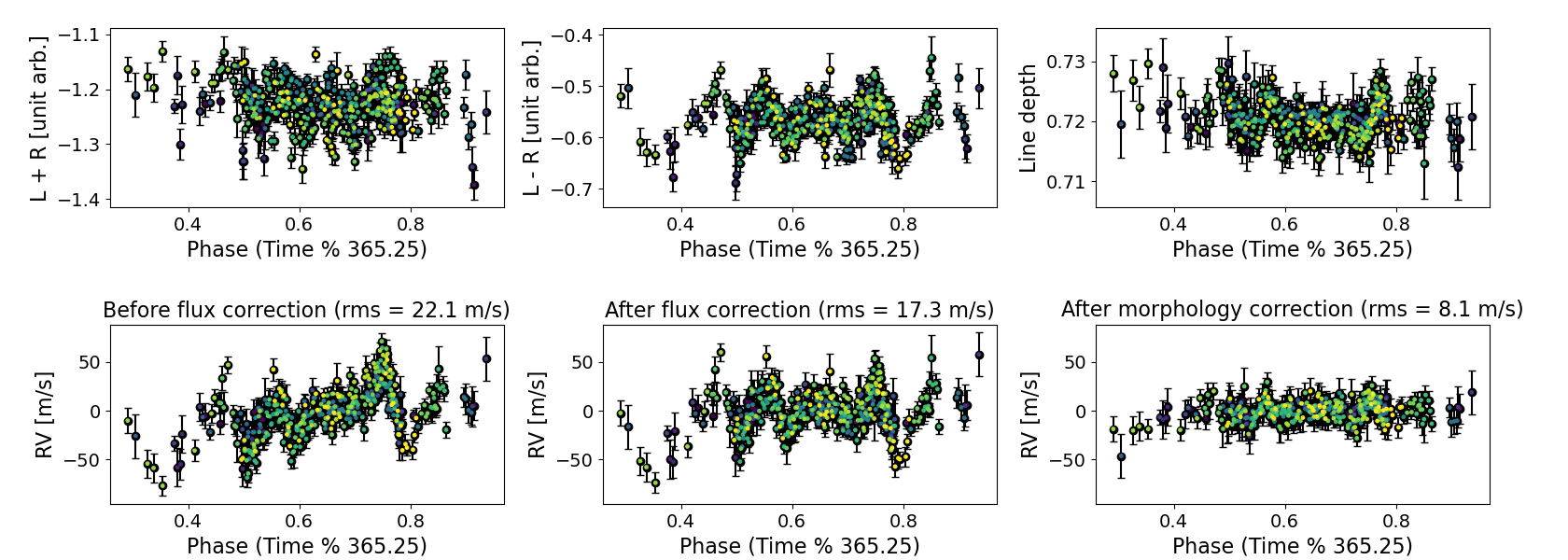}
	\caption{Morphological corrections performed by YARARA on HARPS observations of HD10700 for the 3849 \ang{} line contaminated by a ghost of the science fiber. The time dimension was phase-folded at 1-year. \textbf{Top left:} L+R area deviation by respect to the Doppler shift calibration curve (see main text for the precise definition), therefore similar to its EW. \textbf{Top middle:} L-R area deviation by respect to the Doppler shift calibration curve. \textbf{Top right:} Line depth of the line computed by a parabola fitting on the line core. \textbf{Bottom left:} LBL RV of the line before any correction. \textbf{Bottom middle:} LBL RV after YARARA flux corrections. The stellar lines present in the ghost are still visible. \textbf{Bottom right:} LBL RV after multi-linear decorrelation with the three morphological proxies. The 8.1 m/s rms of the RV residuals is compatible with the median of the RV uncertainties.}
	\label{FigMorpho}
\end{figure*}

%\st{In order to produce such masks, it is important to be accurate on the line depth since these latters directly provide in one hand the weights in the CCF} \citep{Pepe(2002)} \st{and in the other hand the line profile gradient $\partial f/\partial \lambda$ used in the LBL. Not accounting for it could result in a overweighting of the shallow lines in the blue spectral region. The accuracy is recovered by measuring the deviation of the RASSINE continuum by respect to a stellar template close to the effective temperature of the star. This correction was already mentioned in} \citep{Cretignier(2020b)} \st{due to its easy implementation in a post-processing pipeline and we decided to develop it for YARARA. An example of a M1.5V stellar master spectrum corrected for the continuum opacity} (Fig.~\ref{FigContiGJ654}) \st{as well as the detailed description of the process is presented in} Appendix.\ref{app:continuum}. %The main idea consist in measuring the deviation of the continuum with the template spectrum, before to fit a smooth model to link the points. 

\begin{figure*}[h]
	\includegraphics[width=18cm]{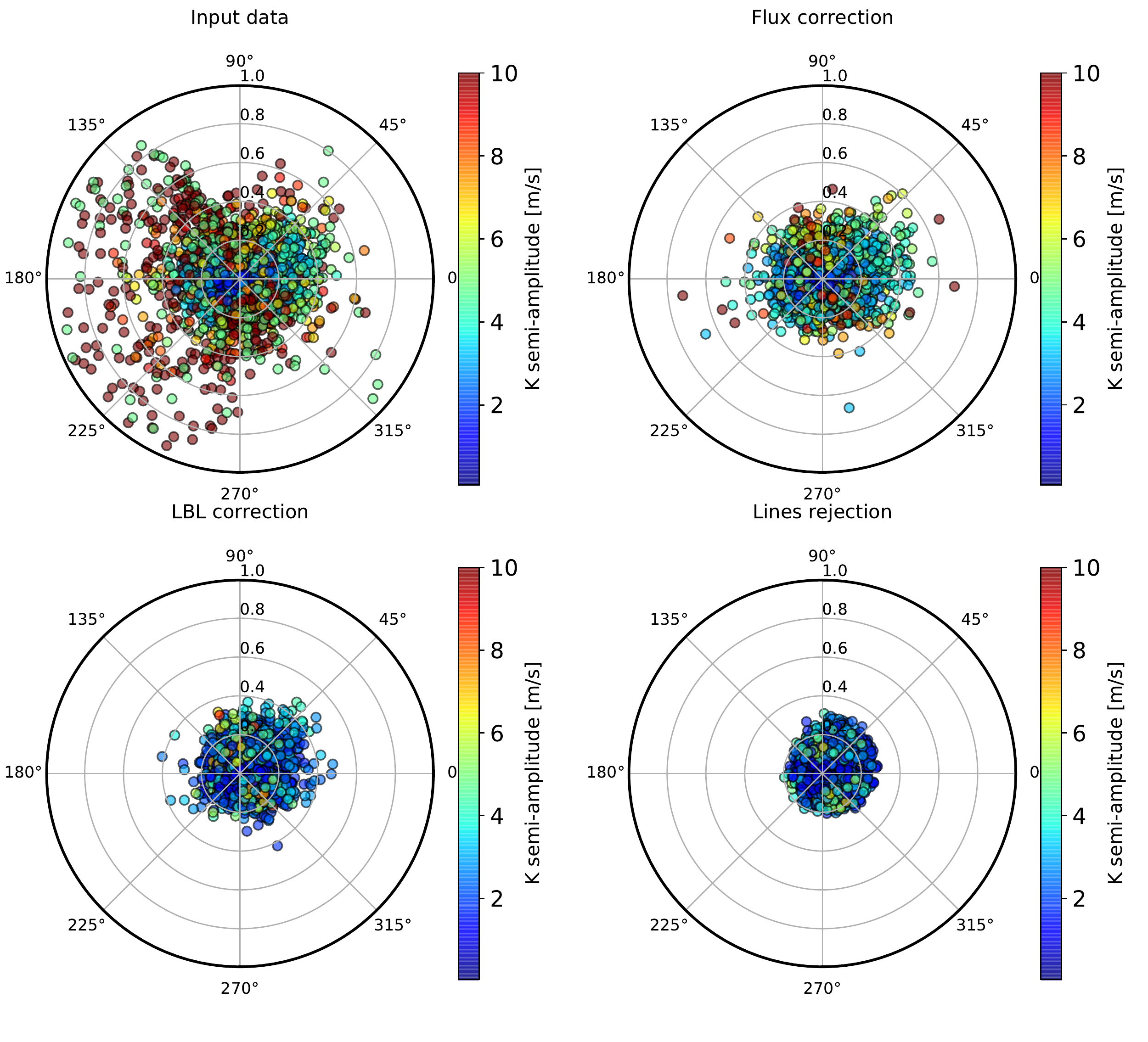}
	\caption{Polar periodogram representation of LBL RV time-series for HD10700. The polar periodogram is focused at 1-year and represent the Pearson coefficient with 1-year sinusoid as radial coordinates (see the main text) and the phase of the signal as polar angle. Among the 3770 stellar lines, the number of lines presenting a Pearson coefficient higher than 0.4 is respectively from top left to bottom right of: 339, 59, 15 and 0. \textbf{Top Left:} Input dataset without any YARARA correction. \textbf{Top right:} Dataset after YARARA flux correction. \textbf{Bottom left:} Dataset after the LBL morphological corrections.  \textbf{Bottom right:} Dataset after lines with a significant 1-year power are rejected. Most of the 1-year have been cleaned out.}
	\label{FigPolar}
\end{figure*}

\subsection{LBL morphological corrections\label{sec:lbl}}

All the previous corrections managed at some level to remove efficiently the different observed contaminations. In some cases, however, we noticed that some residual signals were still present. Those are due to 1) degeneracies between the recipes which typically occurs when several contaminations are located at the same wavelength position, to 2) more complex patterns like the ones created when ghosts contaminate stellar lines (see Sect.~\ref{sec:ghost}) and to 3) regions not flagged by our static products. Unlike a Doppler shift, those residuals will change the shape of spectral lines with time, which can be tracked and corrected for using a few metrics describing line morphology.

%\st{Hopefully, if these contaminations outreach the flux level precision, they can be distinguish from the effect of a pure Doppler shift by a few metrics based on line profile morphology variations.}

We extracted three morphological parameters to measure a change in the line profile not related to a Doppler shift. We first separate the left and right wing of a spectral line with respect to its core defined as the local minimum of the line, and then measure the total flux on the left (L) and right (R) wings. We then build the L+R and L-R metrics to account for asymmetric variations. Our third parameter is the depth of the stellar line, fitted by a parabola on the line core with a window of $\pm 2$ kms.

Although a Doppler shift does not affect the line depth, it does affect L and R, and thus our metrics L+R and L-R. To correct for Doppler shift variation, we built a calibration curve for each stellar line in our line selection. To do so, we shifted all the spectral line with values ranging from -100 to 100 m/s by step of 0.5 m/s and measured for each of them how L-R and L+R behaved. We then linearly interpolated each relation on a finest grid of 1 cm/s. Once we have those calibration curves, the L+R and L-R metrics for a given spectral line can be obtained by 1) deriving its LBL RV, 2) read on its calibration curve the L-R and L+R values for this RV and 3) measuring the difference between the measured and theoretical values. By doing this calibration and correction, we ensure that our morphological proxies L+R and L-R are planetary-free. We demonstrated in Sect.~\ref{sec:results} that neither known planets, nor injected ones, disappeared at the end of our pipeline, which is a first validation of the method.

%\st{define the left window area (L) and right window area (R) which are the two windows on both sides of the line core. If a line is undergoing a Doppler shift, the area in both windows will naturally change, we therefore build a Doppler calibration curve for each stellar line in our mask. The calibration for a line was build as followed. The line profile of the master spectrum is first extracted, before being shifted between -100 and 100 m/s by step of 0.5 m/s. For each shifted line profile, the L and R area are computed. The Doppler calibration curves of L-R and L+R as a function of the RV shift were then linearly interpolated on a finest grid of 1 cm/s. Measuring the L-R or L+R parameter of a stellar line consists in three steps : 1) deriving its LBL RV, 2) read on its calibration curve the L-R and L+R value for this RV value, 3) measuring the difference between its measured and theoretical Dopplerian value. By doing so, we ensure that our morphological proxies are planetary-free. We demonstrated in} Sect.~\ref{sec:results} \st{that neither known planets, nor injected ones, disappeared at the end of our pipeline, which validates this method. The depth of the stellar line, fitted by a parabola on the line core, is finally used as this morphological parameter is obviously planetary-free.}

Those morphological parameters, phase-folded with a period of 1-year, are represented in Fig.~\ref{FigMorpho} for the stellar line at 3849 \ang{} in HD10700 HARPS observations. This line was clearly showing residuals after ghost correction in Fig.~\ref{FigGhost}. %By looking the line-depth time-series, we can identify two peaks (the first at phase 0.48 and the second at 0.78) which indicate that at least two stellar ghost lines are crossing over the stellar line. 
By observing the LBL RV, after flux correction (bottom center panel), several "mountain peaks" are clear visible, a signature coherent with several flux anomalies crossing the stellar line. Because this variation is not equivalent to a pure Doppler shift, the same pattern can be observed in the L-R proxy (top center panel). By linearly decorrelating the LBL with the three morphological proxies, the RV contamination still remaining after YARARA flux correction is highly reduced. We note that for this line, the RV rms is mitigated from 22.1 to 17.3 m/s after YARARA flux correction, and down to 8.1 m/s after LBL morphological correction. %\st{As a remark, it is interesting to note that a spot on a stellar surface would precisely produce the same signatures} \citep[e.g.]{Dumusque(2014)} \st{ which show that this method of correction for line is very promising to mitigate stellar activity on fast rotating stars. However, it is not clear if, for small spots, the flux anomaly would be sufficient to be measured on each individual line and lines stacking could be necessary.} \xav{TOO MUCH SPECULATIVE AND IN THE END< VERY SIMILAR TO A BIS + FWHM CORRECTION AT THE CCF LEVEL, AS A SPOT WILL LIKELY AFFECT ALL LINES IN THE SAME WAY}.

The same morphological correction was systematically applied for all of the 3770 stellar lines in our line selection for HD10700. We displayed in Fig.~\ref{FigPolar} how such a correction is helpful to correct for 1-year period contaminations. We fitted on each LBL RV a 1-year sinusoid and studied the correlation between the time-series and the best-fitted model. In the different panels of this figure, we show for each line, as radial coordinate the weighted Pearson correlation coefficient between the LBL RVs and the fitted sinusoid and as polar coordinate the phase of the fitted sinusoid. The color represents the semi-amplitude of the fitted sinusoid. We defined a line as significantly affected if its power exceeds a false alarm probability (FAP) level of 1\% at 365.25 days. We see in Fig.~\ref{FigPolar} that before any correction is applied (top left panel), 1071 stellar lines exhibit a significant 1-year power, some of them with semi-amplitudes larger than 10 m/s. The third quartile (Q3) of the amplitude's distribution is 9.1 m/s. Despite the large number of lines contaminated, it is important to highlight that since the phases between all the signals are not coherent, the final amplitude of the contamination is roughly of 1 m/s. Those lines can be contaminated by any effect fixed in the BERV rest-frame, in our case tellurics, stitchings, ghosts or Thorium-Argon contaminations. Those contaminations should be significantly mitigated after YARARA flux correction (top right panel), which is indeed the case as only 428 stellar lines still present a significant 1-year signal after flux correction, and Q3 is reduced down to 5.36 m/s. After morphological LBL correction, the number of significantly affected lines drops to 227 and Q3 to 3.2 m/s. Fig.~\ref{FigPolar} therefore highlights that YARARA, through flux and line morphology corrections, is able to strongly mitigate instrumental systematics with a one-year period.

%\st{To obtain such plots we fitted a 1-year sinus on each LBL RVs, and computed the weighted Pearson coefficient between the best fit and the time-series} \xav{I DO NOT LIKE THE COMPARISON WITH THE PERIODOGRAM, AS TH EPERIODOGRAM GIVES YOU THE INFORMATION FOR ALL PERIODS AT ONCE< WHICH IS NOT THE CASE HERE. IT DOES NOT MEAN THAT THIS REPRESENTATION IS NOT INTERETING :)} \st{providing a metric closely related to the classical 1-year power of a Generalised Lomb-Scargle (GLS) periodogram. The Pearson coefficient is represented on the radial axis and the phase of the fitted sinus as the polar angle. The Pearson coefficient appears as a good radial variable since its value ranges between 0 and 1 by construction. This representation  allows to represent thousands of periodogram power simultaneously for a determined period which might be useful for specific cases as here for 1-year power. We will define a line with significant 1-year power if its power exceed a 0.1\% false alarm probability (FAP) level at 365.25 days. }

Those 227 still affected spectral lines are so far not understood. After investigations, no particular instrumental systematics are expected on the detector at their positions. Also, since their contamination is not corrected for by our morphological correction, it indicates that the flux contamination is too low to be detected by our morphological proxies. It is further confirmed by the observation that such remaining 1-year lines are not found after flux corrections for HD215152 for which the median S/N of the spectra time-series (S/N$\sim$120) is ten times lower than for HD10700 (S/N$\sim$1000). This line rejection step is performed by the pipeline for all the stars, but only stars with exquisite S/N like HD10700 will have some lines rejected, generally less than a hundred lines.

We note that when fitting for ghosts localisation on the high S/N master FLAT, some barely visible ghosts were excluded as considered as negligible (see Appendix \ref{app:modeling}). Those could therefore be responsible for the remaining affected spectral lines. To remove from our final selection part of those 227 contaminated lines, we proceed as followed. All the lines presenting a power at 1-year higher than a FAP level of 1\% are first removed and the average RV signal $RV_m$ of the complementary lines' selection is computed. This $RV_m$ signal was then subtracted in each individual LBL RVs, in order to remove any real planetary signal if any, before redetermining the power at 1-year for each individual line. The motivation behind subtracting $RV_m$ is to prevent absorbing any real planetary signal appearing at a period of 1-year or corresponding harmonics (see the planetary injection simulation performed in Sect.~\ref{sec:hd10700}). Lines with a 1-year power higher than 1\% FAP level are removed such that only 103 lines still present significant 1-year power for HD10700 and Q3 is decreased down to 2.5 m/s (bottom right panel). 

An explanation for that remaining 1-year signal is the inaccurate wavelength solution due to the sparse location of Thorium-Argon lines and their uncertainties on their absolute wavelength values. In \citet{Coffinet(2019)}, the authors showed that echelle-order wavelength solution by polynomial fit were inaccurate by dozens m/s compared to the wavelength solution obtained with a laser frequency comb. The inaccurate wavelength solution will at term introduce systematics at 1-year of small amplitudes on LBL RVs since the BERV will move the stellar lines on this inaccurate wavelength solution. Opposite to ghosts, all the lines will be in phase with the BERV.  This seems to be observed on the bottom right panel of Fig.~\ref{FigPolar} by an elongation of the cloud along the $45^{\circ}$ direction, which is almost consistent with the BERV phase of $30^{\circ}$ that we can obtain by doing a similar analysis not shown here. Due to the small amplitudes of the signals expected ($<5$ m/s from the line-by-line RVs), it is rather hard to confirm this hypothesis. %We note that the ESPRESSO DRS has been recently adapted for HARPS which has perharps resolve this issue since a new method was developed to fit the wavelength solution on the full detector and using the FP.

More sophisticated LBL RV corrections could be developed from this point, but those will be described in a forthcoming paper dedicated to stellar activity correction. For now, this stage is representing the ending point of our pipeline. Its application on several HARPS targets is presented in the next section.

\section{Results} \label{sec:results}

In this section, we analysed the data of three stars intensively observed with HARPS, to test the performances of YARARA. We first looked at the data of \object{HD10700} in Sect.~\ref{sec:hd10700}, as it appears to be the quietest star in the entire HARPS sample, with a RV rms close to 1 m/s over more than a decade. We compared the RVs obtained after YARARA processing, with the ones from the classical HARPS DRS and the SERVAL post-processing pipeline \citep[][]{Zechmeister(2020), Trifonov(2020)}. As this star does not show any significant planetary signal, we also injected a fake planetary signal to demonstrate that YARARA processing was not altering planetary signals. We then analysed \object{HD215152} in Sect.~\ref{sec:hd215152}, to demonstrate that YARARA strongly mitigate the impact of yearly signals induced in this dataset mainly by detector stitchings. Finally, we studied \object{HD10180} in Sect.~\ref{sec:hd10180} that is known to harbor 6 planets. We show that all the planets are recovered with high confidence after YARARA processing, and the obtain orbital parameters are similar with the published values, however, with some interesting differences. %\commick{I would like to mention the change of the mass since as you know mass accuracy is really important even though really difficult to guaranty. By the way orbital parameters are not compatible for the planet in question.}
%\st{we demonstrate that any of the 6 confirmed planets are absorbed, but the mass of HD10180 $f$ was slightly overestimated by 18\% due to the 1-year contamination. The orbital parameters after YARARA correction for the outermost planets are also slightly different.} %Finally, we tested the time-transit accuracy $T_0$ of our new RV time-series to see if better agreements were found with transiting planets on HD136352 (Sect.~\ref{sec:hd136352}).   

\begin{figure*}[h]
	\includegraphics[width=18.5cm]{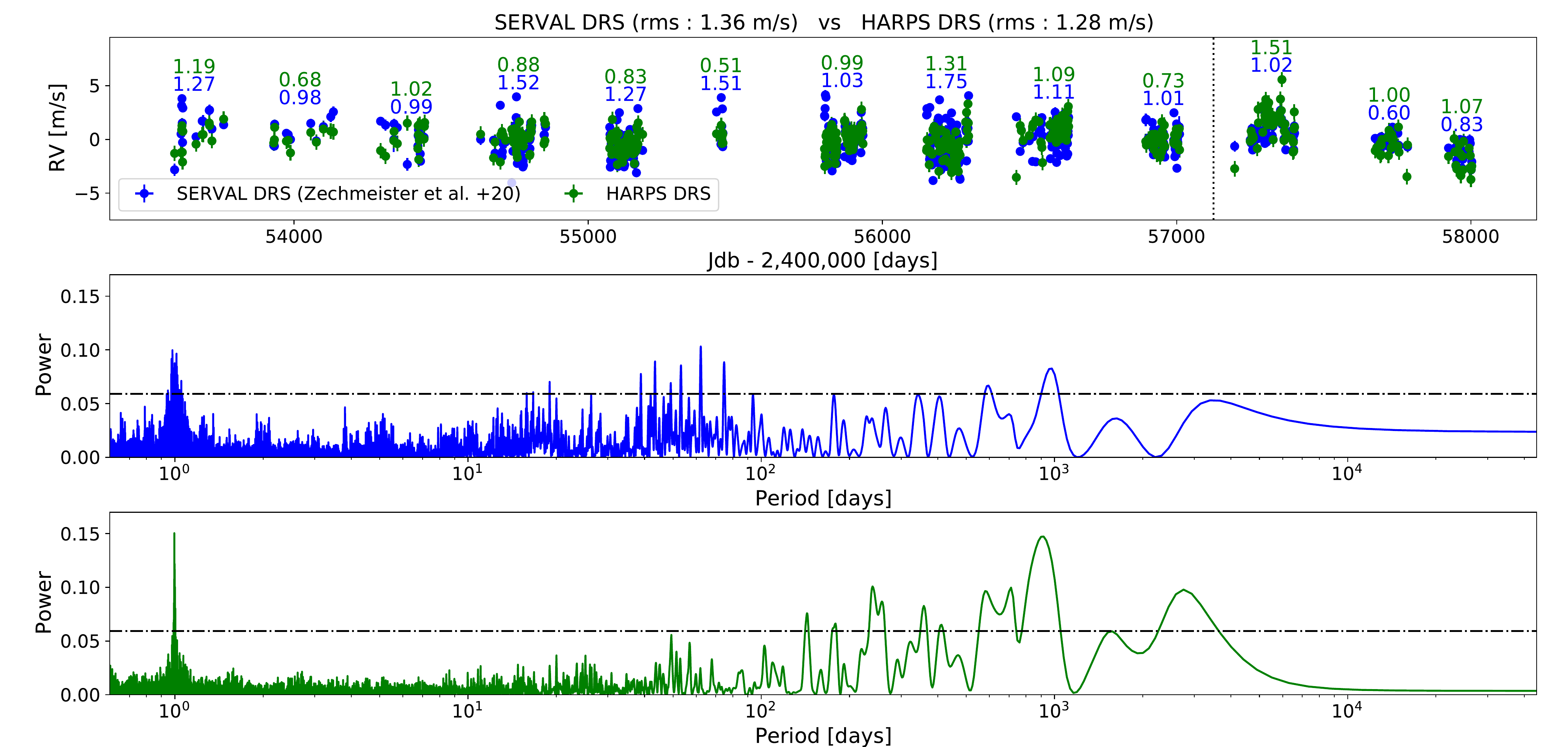}
	\caption{\textbf{First row:} Comparison between SERVAL post-processing pipeline \citep[without night-to-night offset correction,][]{Trifonov(2020),Zechmeister(2020)}  time-series (blue dots) and HARPS DRS (green dots) for 13 years of HD10700 HARPS observations. The weighted rms are indicated for each observational season as well as for the full time-series in the title. HARPS DRS time-series presents a significant lower rms in each season before the fiber upgrade (vertical dotted line), whereas the opposite conclusion holds for SERVAL after it. \textbf{Second row:} GLS periodogram of the SERVAL time-series. The 1\% FAP level is indicated by the dotted-dashed line. \textbf{Third row:} Same as middle row for HARPS DRS extracted time-series.}
	\label{FigCompServalDrs}

	\includegraphics[width=18.5cm]{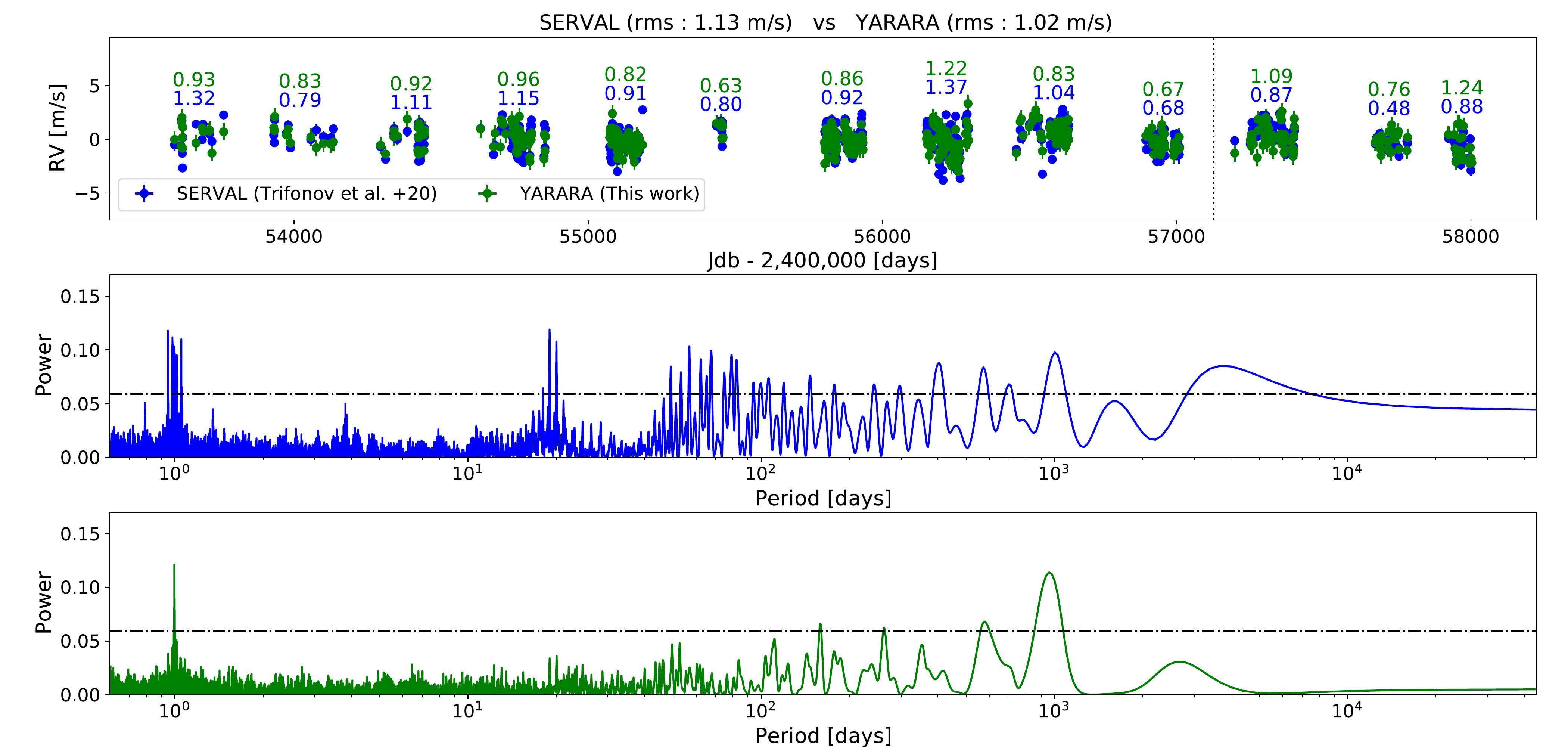}
	\caption{Same comparison as Fig.~\ref{FigCompServalDrs}, except that the comparison is now between YARARA time-series (green dots) and SERVAL post-processing pipeline (with night-to-night offset correction, blue dots) presented in \citet{Trifonov(2020)}. The rms is lower with YARARA for all the observational seasons before the fiber upgrade. The periodogram also present a cleaner aspect. }
	\label{FigCompServal}
\end{figure*}

\subsection{HD10700 \label{sec:hd10700}}

We used HD10700 as a star of reference to perform pipeline performance comparison and a planetary injection test. This star is known to be a particular inactive star maybe in a state similar to the Maunder minimum for the Sun \citep{Lovis(2011)}. On 13 years of HARPS observations, the RV rms is observed to be 1.27 m/s, which is an impressive proof of stellar and instrumental stability. Moreover, this system has been intensively followed by HARPS and contains no confirmed planet but only possible candidates with RV amplitudes lower than 1 m/s, even if no agreement was found between different analyses \citep{Pepe(2011b),Tuomi(2013), Feng(2017)}.

\begin{figure*}[h]
	\includegraphics[width=18cm]{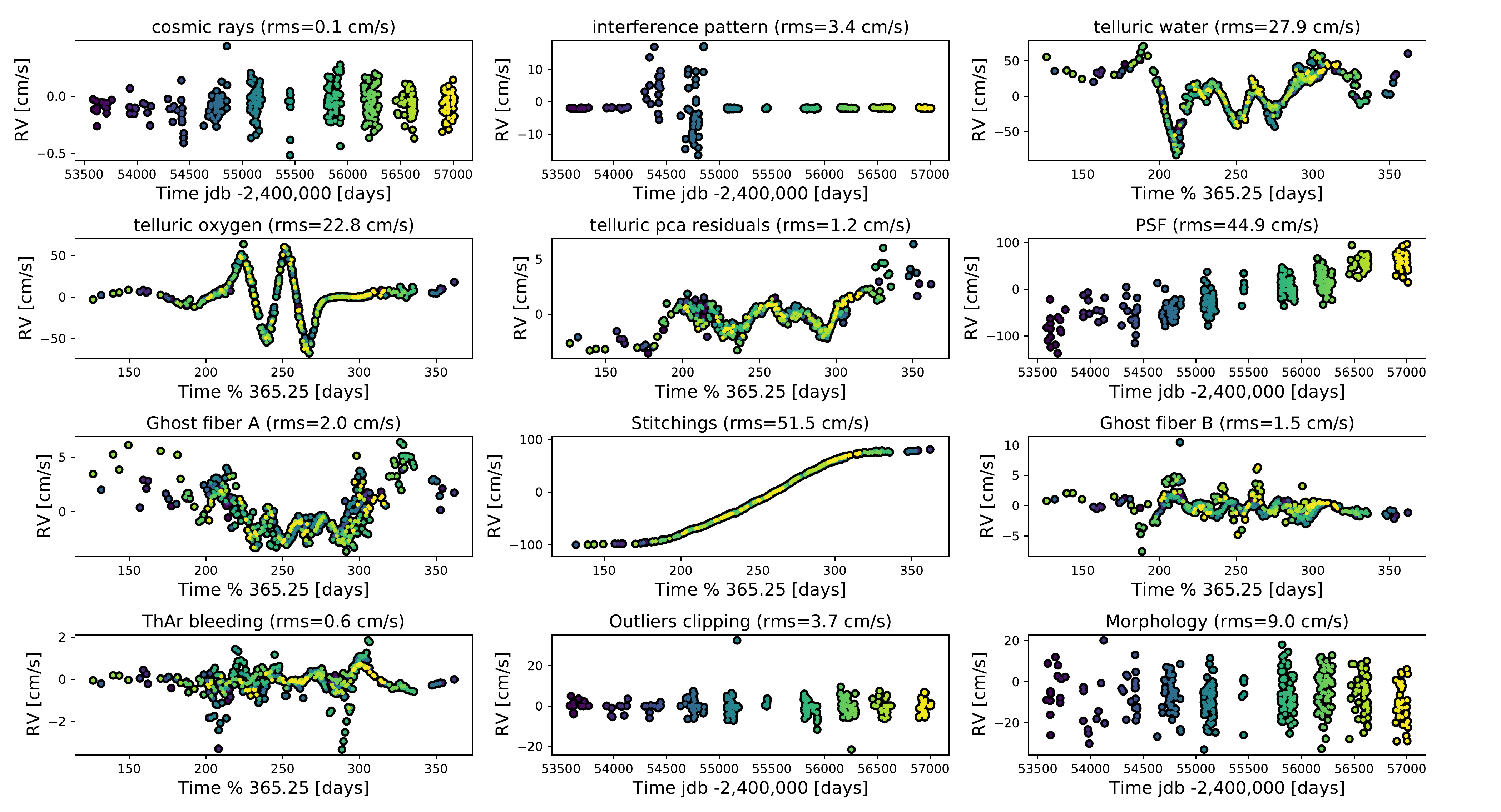}
	\caption{Decomposition of the individuals components corrected in YARARA for HD10700 on HARPS. Each RV component was obtained as in Fig.~\ref{FigFwhm} by computing the RVs before and after the flux corrections of a dedicated YARARA recipe. The RVs are obtained from the tailored lines' selection. Systematics known to be related to 1-year systematics are time-folded. Note that systematics can be folded at 1-year but are not described by sinusoid, explaining why their power can leak afterwards on others frequencies in periodogram. Despite the moderate rms of some components, the peak-to-peak of most of them is above 10 cm/s.}
	\label{FigComponents}
\end{figure*}

We first compared the publicly available RVs obtained with the SERVAL post-processing pipeline \citep{Trifonov(2020),Zechmeister(2020)} with the ones obtained by the HARPS DRS in Fig.~\ref{FigCompServalDrs}. We note that in the publicly available SERVAL RVs, two different products are available. RVs computed using a stellar template matching technique, and the same RVs but corrected for night-to-night calibration offsets \citep[e.g.][]{Dumusque(2018)} measured on the time-series of several quiet stars. To have a fair comparison with the DRS RVs, we considered the SERVAL RVs without night-to-night offset correction. We see that for HD10700, SERVAL does not performs as well as the DRS, with a RV rms over the entire time-series of 1.36 compared to 1.27 m/s, and the RV rms for each season being higher than for the DRS for the data before the change of the HARPS fibres in 2015 \citep[][]{LoCurto(2015)}. SERVAL is performing best after the fiber upgrade of the instrument. In addition, while SERVAL seems to slightly remove some signals with periods longer than a hundred days, it seems that more signals are present between 40 and 100 days, which is worrisome. It is clear from the \citet{Trifonov(2020)} paper that SERVAL performs better than the DRS for M-dwarfs, as the stellar template-matching approach used by SERVAl is known to work better for these stars \citep[][]{Anglada(2012)}. However, the \citet{Trifonov(2020)} paper does not asses the performance of the code with respect to spectral type, and this example shows that for G dwarfs, SERVAL does not necessarily outperform the DRS.

We then compared the SERVAL RVs with the ones obtained by YARARA. As YARARA works with nightly-binned spectra, and thus RVs, we nightly-binned the SERVAL data to perform the comparison. Two nights were clearly outliers in SERVAL reprocessed data, so we removed them in both datasets. In this case, to perform a fair comparison, we took the SERVAL RVs with the night-to-night offset correction applied. As seen in Fig.~\ref{FigCompServal}, correcting for the night-to-night offset significantly improve the SERVAL RVs in term of rms (1.13 m/s), however, the forest of peaks between 40 and 100 days is still present. In comparison, the global RV rms is reduced down to 1.02 m/s after YARARA processing, which is equivalent to a quadratic difference improvement of 0.77 m/s compared to the HARPS DRS. We highlight in Fig.~\ref{FigComponents} for HD10700, the different contaminations corrected by YARARA to show the relative strength of each component relative to the others. We can see that most of the contaminations are peak-to-peak larger than 10 cm/s. The dominant systematics is produced by the stitching \citep{Dumusque(2015)} with 1 m/s peak-to-peak variation. %Fitting signals more significant than 1\% FAP level on YARARA post-processed RVs leads to a final rms of 0.90 m/s on the full lifetime of the instrument. %In addition, except for the signals around 600 and 1000 days, which are also present in the SERVAL and DRS RVs, no other signals are significant at the 1\% FAP level in the periodogram of YARARA RVs. 

By comparing Fig.~\ref{FigCompServalDrs} and Fig.~\ref{FigCompServal}, we can see that YARARA provides better individual season RV rms compared to the DRS for only half of the observational seasons, which highlights the difficulty to significantly improve the RV below 1 m/s on HARPS. Despite those differences, by looking at the global RV rms and the periodograms of the different time-series, YARARA seems to clearly outperform SERVAL and the DRS.

%\st{As a remark the improvement brought by YARARA will highly depends on the relative position of the contaminations by respect to the stellar lines and thus more notable improvement can be obtained for others systems} (see HD215152 in Sect.~\ref{sec:hd215152}). \st{Nevertheless, the rms is not the most relevant metric to judge the RV precision. A more notable approach is to compare how periodogram compared to each other and, at this respect, YARARA presents the cleanest landscape.}
%\st{Such method is assuming that others signal as stellar activity or unknown planets will be averaged out during the process which is a strong hypothesis.}

%\begin{figure*}[h]
%	\includegraphics[width=18.5cm]{Injected_planet_HD10700_k3_e1.png}
%	\caption{Planetary injection in the HD10700 dataset. Comparison between the injected Keplerian parameters (black dotted line) and the recovered ones obtained after fittting a 3-Keplerian model and linear drift. Each column represent one of the Keplerian parameter. No planetary absorption is observed after YARARA processing (orange and green marker), since similar values than those without correction are found (blue markers). \textbf{First row:} First injected planet at 37.87 days all parameters are recovered at 2-sigma. \textbf{Second row:} Second injected planet at 122.42 days. All the parameters are recovered at 2-sigma, excepted for the periastron angle and ascending node before YARARA correction.}
%	\label{FigKepHD10700}
%\end{figure*}

YARARA seems to perform better in correcting for the different instrumental systematics, and thus give RVs with a smaller rms and that present a cleaner periodogram. However, this would be useless if YARARA processing absorbs planetary signals. To check that this was not the case, we injected into the spectra, before YARARA processing, the signature of two planets in circular orbits. A first one close the second harmonics of 1-year at 122.42 days and with a RV semi-amplitude of $K=3$ m/s, and a second one at 37.87 days with $K=2$ m/s. The first planet is equivalent to planet $f$ in HD10180, and since our analysis of HD10180 (see Setc.~\ref{sec:hd10180}) shows that the semi-amplitude of planet $f$ change significantly after YARARA processing, we wanted to be sure that the recipes dealing with yearly instrumental effects were not perturbing the amplitude of planetary signals at the harmonics of a year.

To inject the planetary signals in HD10700's spectra, we proceeded as follows. First, we ran HD10700's data through YARARA, as described in Sect.~\ref{sec:theory}, and we kept at each stage the contamination map. Secondly, merged 1D corrected spectra are shifted according to the planetary signals, before reintroducing the extracted contaminations. It would be wrong to simply shift the spectra still contaminated by systematics, since the planetary signals would affect the instrument systematics. For example, telluric lines would not be fixed anymore in the BERV rest frame. Once the planets are properly injected in the original spectra, we ran again YARARA to check that the planetary signals are not absorbed.

\begin{figure}[h]
	\includegraphics[width=9cm]{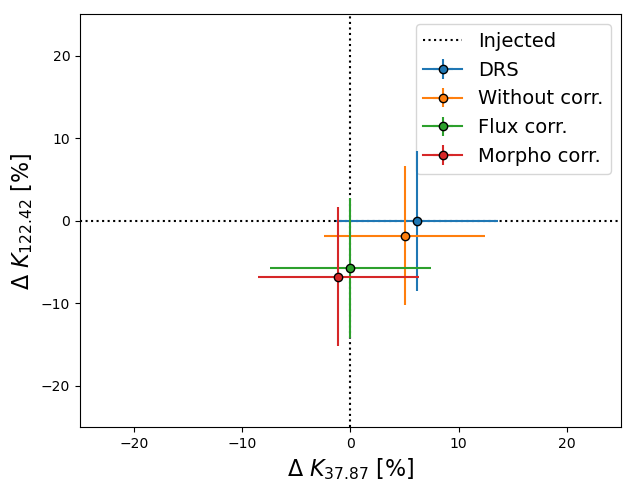}
	\caption{Planetary injection recovery in the HD10700 dataset. Comparison between the injected $K$ semi-amplitudes (black dotted line) and the recovered ones obtained after fitting a 2-sinusoid model for the planets at 37.87 ($K=2$ m/s) and 122.42 days ($K=3$ m/s). No significant absorption of the planetary signals is observed after YARARA processing, neither by flux correction (green markers) nor by morphological correction (red markers), since similar values than those without correction (orange markers) or DRS (blue markers) are found. All the recovered amplitudes are compatible with the injected values considering the uncertainty due to preexisting signals.}
	\label{FigKepHD10700}
\end{figure}

Both injected planets were found to produce the highest peak in the GLS periodogram as expected from the high amplitudes of the planetary signals injected. However, to validate the efficiency of YARARA in recovering planetary signal while mitigating other systematics, we needed to measure at which precision the injected planetary signals were recovered. In Fig.~\ref{FigKepHD10700}, we compare the semi-amplitude $K$ for the best-fitted Keplerian model for the two injected planets in the following RV time-series: 1) DRS (blue marker) 2) before YARARA correction (yellow marker) 3) after YARARA flux correction (green marker) and 4) after YARARA morphological corrections and lines rejection (red marker). There are few differences between the RVs of the DRS and YARARA before correction, justifying their different treatment. YARARA is working with 1D-merged spectra while the DRS starts from 1D-echelle-order spectra. In addition, the color correction is performed differently, the lines selected for the CCF mask are different, and the cross-correlation algorithm is different.

%\xav{IN THE ORIGINAL TEXT, you had also after line rejection. YOU DELETED THIS PART IN THE FIGURE ?}
%\st{1) DRS (black marker) 2) before YARARA correction (blue marker) 3) after YARARA flux correction (yellow marker) and 4) after YARARA morphological corrections (green marker) and 5) after lines rejection (red marker).} 

\begin{figure*}[h]
	\includegraphics[width=18.5cm]{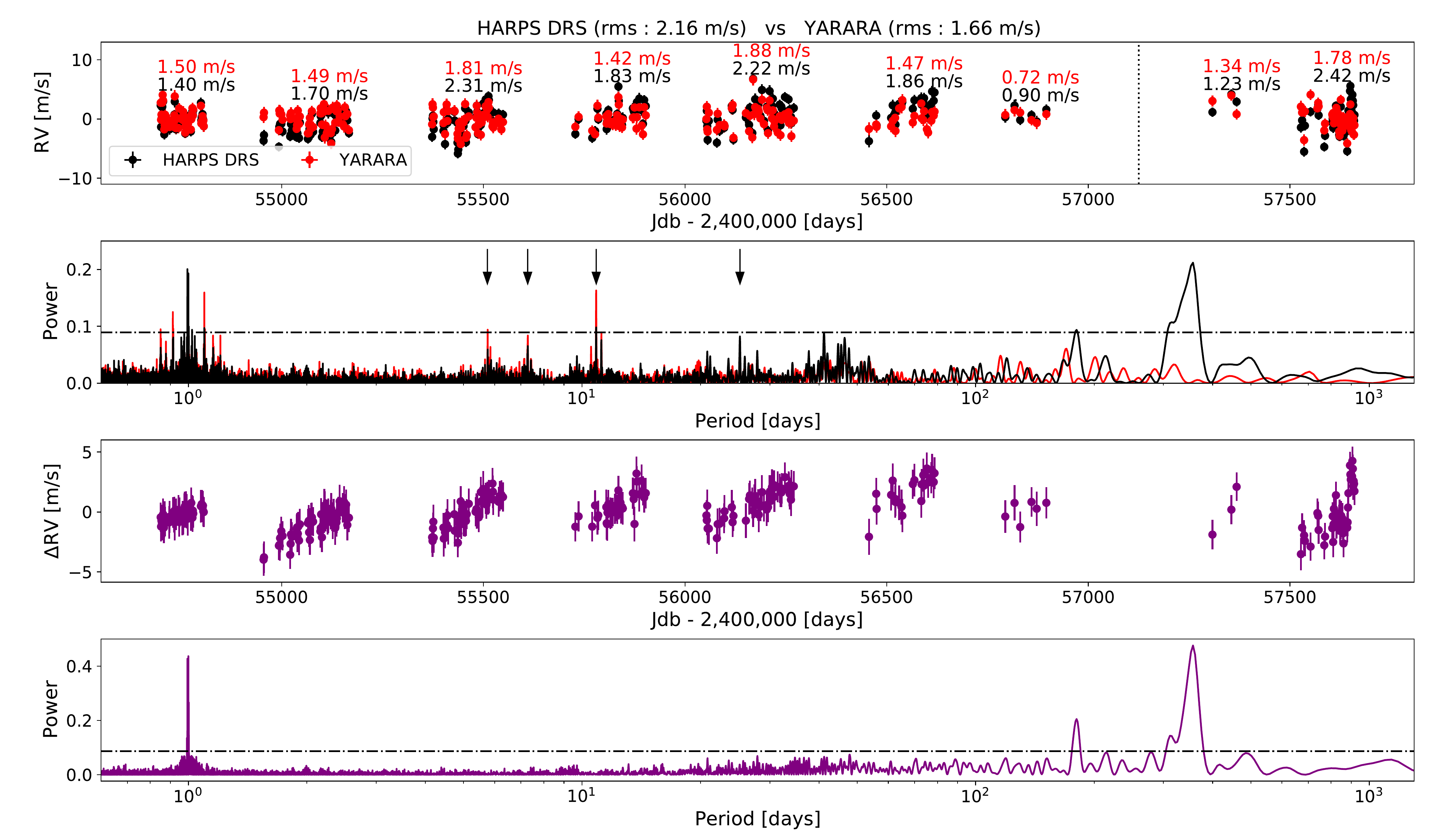}
	\caption{\textbf{First row:} Comparison of the HD215152 RV time-series extracted with the HARPS DRS (black dots) and after YARARA processing (red dots). \textbf{Second row:} GLS periodogram of the previous time-series. 1\% FAP level is indicated as a dotted-dashed line. Exoplanets reported in Delisle(2018) are indicated by black arrows. After YARARA processing, the power at the planetary periods (5.8, 7.3, 10.9 days) are boosted. \textbf{Third row:} RV difference between input and output data. \textbf{Fourth row:} GLS periodogram of the RV difference. The removed contaminations were clearly producing most of the signals at one year and its first harmonic.}
	\label{FigPerioHD215152}
\end{figure*}

To determine an uncertainty on the semi-amplitude that could be due to preexisting power in the time-series, we fitted the same model on the DRS time-series without the planet injected and reported the amplitudes of the signals as error bars. The relative uncertainties compared to the injected amplitudes is of $7\%$. %The jitter is also fitted in the noise model. The fit was performed on the interactive DACE platform according to \citet{Delisle(2016)} and followed by a MCMC analysis with an uniform prior. In Fig.~\ref{FigKepHD10700}, we report the median as well as the $1\sigma$ value of the marginal distributions. 

All the planetary semi-amplitudes recovered using the different datasets are compatible with the injected values. This is expected as the corrections performed by YARARA reduce the rms by 0.29 m/s, whereas the amplitude of the injected signals are an order of magnitude larger, 2 and 3 m/s. We notice that after YARARA correction, we are able to recover the exact amplitude of the injected 37.87-day signal. It is not the case for the 122.42 injected planet for which a value $\sim$6\% smaller is found. This smaller value, still compatible with the injected signal at one sigma, could be explained by the fact that the signal is at a harmonic of a year, and many different corrections in YARARA are mitigating systematics at a year. Thus, there could be small cross-talks between YARARA corrections and planetary signals close to a year or a harmonic of it. However, this test on HD10700 shows that at most 10\% in the semi-amplitude signals at one year or a harmonic of it could be absorbed by YARARA, which is in any case better than leaving perturbing yearly signals in the datasets. From this simulation of planet recovery, it seems that YARARA does not absorb significantly planetary signals in the final RV time-series. Further tests on other stellar systems, presented in the next two sub-sections, will confirm this conclusion.

\subsection{HD215152 \label{sec:hd215152}}

HD215152 is a compact system of four planets in close-in orbits (< 30 days) and close to a resonant configuration \citep{Delisle(2018)}. In that paper, the authors already reported a 1-year signal present in the RV data that was interpreted as mostly induced by the stitching of the HARPS detector. In order to remove this signal, the authors removed stellar lines crossing each stitching from the CCF mask. Therefore, their strategy consisted in masking instead of correcting for the signal. In this section, we demonstrate that by applying YARARA, the perturbing one-year signal is completely mitigated without masking any spectral region, which has the advantage of improving the RV precision of the final dataset.

An interesting feature of YARARA resides in the fact that since the pipeline is applied sequentially, it is possible to quantify the RV perturbation of each contamination. To do so, the difference of the RV time-series before and after the flux correction of each recipe is computed, directly providing the RV counterpart of the contamination. This method was used in Sect.~\ref{sec:activity} to highlight the instrument ageing on HD10700. With YARARA, we confirm that stitchings are the main source of the one-year signal in the RV time-series of HD215152, with a peak-to-peak amplitude of 1 m/s as for HD10700 (see Fig.~\ref{FigComponents}. %\xav{NOT VERY RELEVANT}\st{It is relevant to note that, the intensity of each contaminations will differ for each star depending on its spectral type and its stellar systemic velocity, which will distribute the contaminations at different wavelength positions on the stellar spectrum, therefore producing different configurations of phases (as the incoherent phase configuration reported for HD10700 in} Fig.~\ref{FigMorpho}). \st{We can therefore expect to see star "stitching-dominated", "telluric-contaminated" or "ghost-dominated" even though this later case is less expected since ghosts are mainly found in the blue, a region under-weighted for any spectral type in terms of RV. }

In Fig.~\ref{FigPerioHD215152}, we displayed the RV time-series before and after YARARA corrections. The RV rms is decreased from 2.14 down to 1.57 m/s (equivalent to 1.45 m/s in quadratic difference). The GLS periodogram of each time-serie is showing that the 1-year signal and its first harmonic are considerably reduced whereas the power at planetary periods (5.7, 7.3 and 10.5 days) are boosted. By looking at the RV difference, a clear linear trend can be observed for each season which is induced by the stitching effect \citep[][]{Dumusque(2015)}. The jitter on smaller timescales is produced by the tellurics. 

\subsection{HD10180 \label{sec:hd10180}}

HD10180 is a system composed of at least 6 exoplanets \citep{Lovis(2011b)}, with RV amplitudes larger than 1 m/s. One of them, HD10180 f, has a period of 122.7 days, thus very close from a harmonic of a year. We could therefore think that the 3 m/s amplitude of this planets could be affected by the yearly HARPS detector stitching, as this perturbing effect was only detected a few years later \citep{Dumusque(2015)}. Indeed, due to a complex interplay between real signals and data sampling, a yearly signal such as the one induced by detector stitching can present more power at a harmonic of a year. We therefore applied YARARA on HD10180 HARPS observations to test if contaminating signals with one-year period can affect recovered planetary signals. 

We note that a planetary candidate, HD10180 b, was also announced in \citet{Lovis(2011b)}, but the induced signal with a period of 1.17 days and semi-amplitude of 80 cm/s is not visible in our analysis, probably due to night-binning. Since the same night-binning was performed for all the datasets that we will compare in this section, the existence or not of this ultra-short period planet is not relevant here.

\begin{figure}[tp]
%	\includegraphics[width=9cm]{/Users/cretignier/Documents/LaTeX/LaTeX_yarara/HD10180_l1_periodogram_drs.pdf}
%	\caption{L1 periodogram \citep{Hara(2017)} applied to the HARPS DRS time-series of HD10180. The six exoplanets in the stellar system \citep{Lovis(2011b)} are recovered at high certainty as highlighted by the FAP level detection. }
%	\label{FigL1Drs}

	\includegraphics[width=9cm]{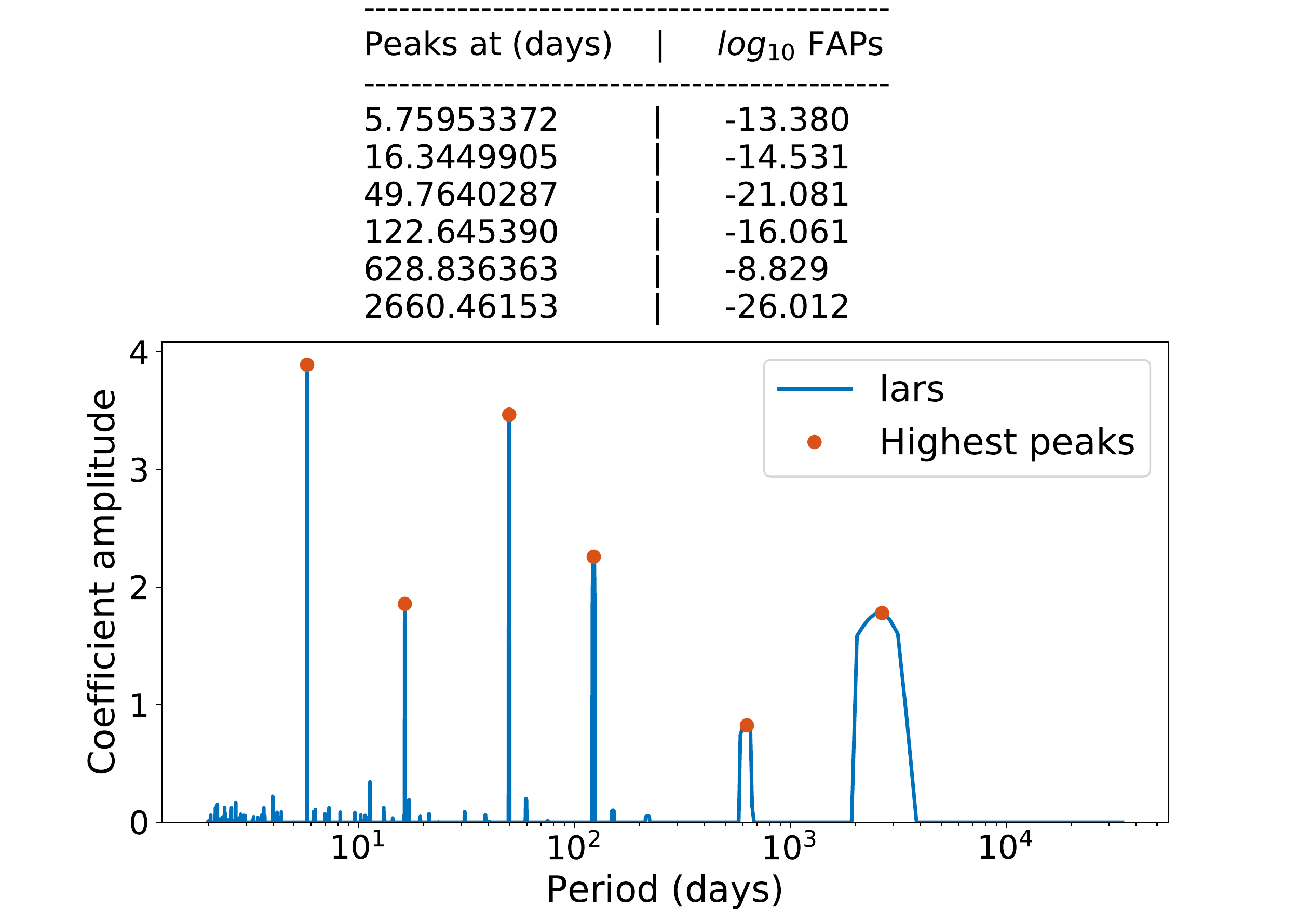}
	\caption{L1 periodogram \citep{Hara(2017)} applied to the YARARA post-processed time-series of HD10180. The six exoplanets in this stellar system \citep{Lovis(2011b)} are recovered at high significance as highlighted by the high FAP values.}
		%Same as Fig.~\ref{FigL1Drs} on the YARARA time-series of HD10180. Any of the six exoplanets disappeared after YARARA post-processing and most detection levels are even higher. The only exception is for the 123-day planet for which the FAP level is reduced, related to 1-year systematics correction. }
	\label{FigL1Yarara}
\end{figure}

\begin{figure*}[h]
	\includegraphics[width=18.5cm]{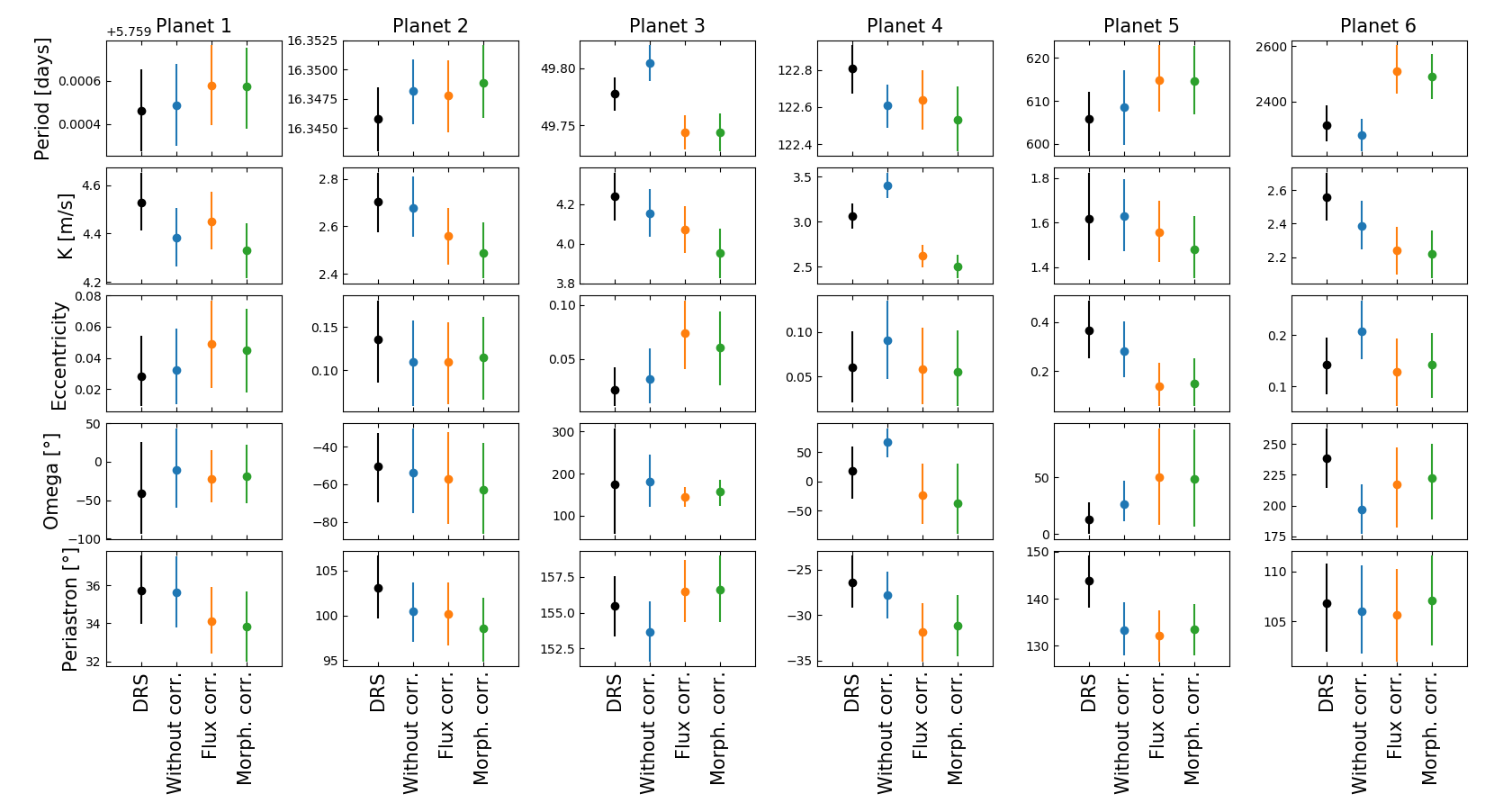}
	\caption{Marginal distributions of the orbital parameters for the 6 exoplanets in the HD10180 dataset obtained with an MCMC sampler. The different datasets analysed are encoded by different colors: the DRS  (black dots), our reduction before flux correction (blue dots), after flux correction (orange dots) and after morphological correction (green dots). Each row represents one of the orbital parameters, each column a planet.}
	\label{FigKepHD10180}
\end{figure*}

After YARARA processing, the 6 exoplanets are well recovered, as we can see from the l1-periodogram\footnote{The python code is freely accessible on GitHub : \url{https://github.com/nathanchara/l1periodogram}}  \citep{Hara(2017)} in Fig.~\ref{FigL1Yarara}.
This is another example that shows that YARARA is not absorbing planetary signals.%The power of the 122-day planets  presents a smaller detection level. This behaviour is well expected if part of the 122-days signal was induced by 1-year contaminations producing an higher amplitude at this period. By correcting those contaminations, the $K$ amplitude becomes lower which decreases the significance of the signal. We demonstrated here after that it is indeed the case by fitting the Keplerians solutions on both datasets.
As done for the planetary signal recovery test performed on HD10700 (see Sect.~\ref{sec:hd10700}), we compared in Fig.~\ref{FigKepHD10180} the orbital parameters of the six planets obtained with the data from the DRS and with the data at different step of YARARA's processing. Those parameters, with their respective error bars were obtained using an MCMC sampler.
In contrast to HD10700, no line was rejected by the 1-year criterion (Sect.~\ref{sec:lbl}), which is expected since the LBL precision is lower for HD10180 due to spectra with lower S/N (median S/N$\sim$160).

When comparing the recovered planetary orbital parameters, we do not observe any significant difference for the three inner-most exoplanets. However, for the 122-day planet, its amplitude decreases by 0.55 m/s after the full YARARA processing compared to the DRS. This 18\% drop is not compatible with the uncertainties. As a comparison, the amplitude of the planet injected at the same period and amplitude on HD10700 (see Sect.~\ref{sec:hd10700}) was only reduced by 7\% after morphological corrections. Therefore, the decrease observed here seems to be real and is likely explained by the interference of the 122-day planetary signal with one-year systematics. This explanation is also favored since in the YARARA input dataset, the planetary signal has an amplitude of 3.4 m/s, which is 15\% higher than the published value. This can be explained by the fact that in YARARA input dataset, all the regions of the stellar spectra are taken into account to measure the RV, even the parts strongly affected by tellurics, which is not the case for the DRS as the mask used for cross-correlation exclude regions of strong telluric contamination. This example of planet HD10180 f demonstrates how some yearly contaminations can interfere with a planetary signal and bias the recovered orbital parameters. The minimum mass of HD10180 $f$ should therefore be updated from $23.6\pm1.3 M_{\oplus}$ down to $19.4\pm1.2 M_{\oplus}$.

We can see other significant differences for HD10180 $g$ and $h$. Using DRS data, the 600-days planet presents a moderate eccentric orbit $(e=0.37\pm0.13)$, which motivated \citet{Lovis(2011b)} to fix the eccentricity of the planet to 0 in their model. After YARARA processing, the period shifts to 615 days and the eccentricity drops to $e=0.15\pm0.10$. Even if we cannot assess with certainty which configuration is the most favored without performing dynamical stability simulations, it seems reasonable that the solution with the smallest eccentricity is favored in this highly populated system. Regarding HD10180 $h$, the period increases to $P = 2500$ days after YARARA processing, and the mass decreases from $52.2\pm3.3$ to $46.3\pm3.4 M_{\oplus}$ \citep[$65.3\pm4.6 M_{\oplus}$ was reported in][]{Lovis(2011b)}.

We do not present here fully updated orbital parameters for HD10180 as this will be the goal of a forthcoming paper. The idea of this section was to give an example that demonstrate that YARARA can help in characterizing better planetary systems by cleaning non-planetary systematics in RV time-series.

\section{Conclusion} \label{Conclusion}

In this paper, we present YARARA, a new post-processing pipeline for deriving precise RV measurements from high-resolution spectra. The purpose of this pipeline is to provide better RV precision by correcting the spectra time-series from different kind of non-planetary systematics using a data-driven approach. The workflow of the code can be described as the following. All spectra in a time-series are first normalised by RASSINE \citep{Cretignier(2020b)}. Then a master spectrum, free as possible from systematics is built and all spectra are compared to this master. Regions showing strong flux residuals are cleaned either by modeling the signal using a Principal Component Analysis (PCA) algorithm and then fitting for the first principal components, or by using a physically motivated model. From the obtained cleaned spectra, it is possible to derive RVs that are less contaminated by different instrumental and astrophysical systematics. We also demonstrate that analysing and correcting for the morphological variation of individual spectral lines after correcting for flux anomalies can improve further the RV precision.

YARARA has been designed to remove cosmic rays (Sect.~\ref{sec:cosmics}), interference pattern (Sect.~\ref{sec:fringing}), tellurics (Sect.~\ref{sec:telluric}), PSF variation (Sect.~\ref{sec:activity}), ghosts (Sect.~\ref{sec:ghost}), detector stitching effects (Sect.~\ref{sec:stitching}), reference fibre contamination (Sect.~\ref{sec:thar}) and continuum precision (Sect.~\ref{sec:smooth}). To our knowledge, it is the first time that a post-processing tool for high-resolution spectra corrects for all those systematics at once.
%\st{a work dedicated to ghosts correction on stellar spectra has been investigated. The multi-linear decorrelation being either performed by wise proxies or by principal components vectors fitted on predefined spectral region known to contain the contaminations. Our approach is therefore a data-driven approach consolidated by the presence of model-driven static products. A more advanced correction based on LBL morphology multi-linear decorrelation (Sect.~\ref{sec:lbl}) was also presented and showed really promising results.}

By analysing the HARPS spectra time-series of HD10700 with YARARA, we demonstrate an instrumental stability on the entire lifetime of 1.02 m/s on nightly-binned spectra (Sect.~\ref{sec:hd10700}). Due to the extreme stability of HD10700, we performed a planetary injection test on the spectra time-series and found that planetary signals cannot be absorbed by more than 10\% in semi-amplitude after YARARA processing. Then, by analysing HD215152, we showed that the strong yearly signal induced by detector stitching is completely removed after YARARA processing, while the known planets in this system are not affected (Sect.~\ref{sec:hd215152}). Finally, looking at HD10180, all the six detected planets are recovered with high significance (Sect.~\ref{sec:hd10180}). Besides a significant change in semi-amplitude for planet $f$, which is likely because this planet has a period of 122 days and therefore is affected by the yearly effect of detector stitchings, all the other orbital parameters are compatible between the DRS and the final step of YARARA processing. A small change in the configuration of the outer part of the system is observed.

Since YARARA is correcting for telluric contamination, it is not anymore needed to exclude strong contaminated regions from cross-correlation masks to obtain RVs that are free of telluric contamination. Deriving RVs after YARARA processing can thus increase the S/N and provide a better RV precision. The gain for the visible bandpass is rather small \citep[<10\%][]{Artigau:2014aa}, but could be huge in the NIR. In addition, the correction of all the known systematics at the spectral level by YARARA is essential to perform further line-by-line (LBL) analysis, as contaminations can induce hundreds of m/s variation on individual spectral lines (see for instance the telluric-contaminated line in Fig.3 of \citet{Cretignier(2020a)}). LBL analysis is a promising path to understand further and mitigate stellar activity perturbations \citep[e.g.][]{Davis(2017), Thompson(2017), Wise(2018), Dumusque(2018)}, and thus YARARA can be of great help there.

Further improvement can be made on the pipeline. A contamination not yet investigated is the one induced by the moon reflection \citep[e.g.][]{Roy(2020)}, which could theoretically be corrected for by training a PCA at the position of strong solar lines in the lunar RV rest-frame. However, such contamination is expected to be very small for high S/N spectra of bright stars, which are the main input products for which YARARA will be efficient. Covariance between wavelength columns in spectra time-series could also be accounted for, in particular to reduce the problems caused by some outliers during the multi-linear regressions.

In this paper, we did not present any updated orbital parameters of the systems that we analysed, and solely mentioned the gain that can be achieved on the mass accuracy in certain cases like for HD10180 $f$ with a mass 18\% lower than what was found in \citet{Lovis(2011b)}. An in-depth analysis of HARPS and HARPS-N observations using YARARA will be presented in the future.

Finally, YARARA is flexible and can be applied to the data of different highly stable spectrographs. The code is already implemented to work with HARPS, HARPS-N, CORALIE, CARMENES, EXPRES and ESPRESSO spectra. Since the pipeline needs to construct a contaminated-free master spectrum to highlight the contaminations before performing PCA, this latter is expected to be more efficient for spectrographs working in the visible and for stars with large BERV coverage values. But a priori YARARA could also be employed for near-infrared (NIR) spectra, if combined with a model-driven approach to correct for tellurics.

Such methods of spectra-time series analysis can be employed to larger perspectives than the primary goal of RV precision and could be for instance used in the context of stellar oscillations in order to detect and measure acoustic modes. In YARARA, spectra are usually nightly binned in order to boost the S/N which is the main limitation of the pipeline. Therefore, it is unclear if such tool may be adapted or not to detect the stellar pulsations of main sequence stars. It could however possibly be used in the context of longer and larger amplitude pulsation modes of variable stars like those of Cepheids \citep[i.e.][]{Anderson(2016)}.

\begin{acknowledgements}
We thank the referee for their really helpful and constructive feedbacks. M.C. acknowledges the financial support of the SNSF. X.D is grateful to the Branco-Weiss Fellowship for continuous support. This project has received funding from the European Research Council (ERC) under the European Union's
Horizon 2020 research and innovation program (grant agreement SCORE No.~851555). F.P. greatly acknowledges the support provided by the Swiss National Science Foundation through grant Nr. 184618

This publication makes use of the Data \& Analysis Center for Exoplanets (DACE), which is a facility based at the University of Geneva (CH) dedicated to extrasolar planets data visualisation, exchange and analysis. DACE is a platform of the Swiss National Centre of Competence in Research (NCCR) PlanetS, federating the Swiss expertise in Exoplanet research. The DACE platform is available at https://dace.unige.ch. This work has made use of the VALD database, operated at Uppsala University, the Institute of Astronomy RAS in Moscow, and the University of Vienna. This work has been carried out within the frame of the National Centre for Competence in Research “PlanetS” supported by the Swiss National Science Foundation (SNSF).
\end{acknowledgements}

\newpage

\bibliographystyle{aa}
\bibliography{Cretignier_2020_YARARA}

\begin{appendix}

\section{Telluric lines' location extraction} \label{app:modeling2}

The detailed correction of telluric lines described in  Sect.~\ref{sec:telluric} is performed by fitting a PCA in the terrestrial rest-frame at the telluric lines' location. To do so, the main step required is therefore to select the so-called "locations", or said differently to be able to form a sub-selection of wavelengths $\lambda_j$ on which to fit the PCA.

Theoretically, this step could be done by simply selecting all the wavelengths known to contain telluric lines deeper than a fixed threshold in percent based on specific database (e.g MolecFit \citet{Smette:2015aa}). Such strategy contains three main disadvantages. First, the pipeline would be no more self-sufficient which go against the main philosophy of YARARA, which is to depend as least as possible on static products. This allows for a relatively simple adaptation of YARARA on any spectrograph, independently of their available spectral range. Secondly, since telluric lines change in depth, any selection based on a depth threshold from a fixed template is somewhat arbitrary. Lastly, the main issue of such strategy is to be independent of the S/N of the observations.

By nature, any data-driven method cannot outperform the noise level of the observations. When performing a PCA analysis, it is not sufficient to guaranty that a telluric line is present at a specific wavelength column $\lambda_j$, we also have to guaranty that the S/N of the signal is sufficient, and this cannot be produced by a static product since it will depend on the S/N of the observations. Again, such conditions should be computed directly from the observations.

To know if the S/N of a telluric line is sufficient, we can proceed as follow. It should be noted that PCA are not sensitive to offsets but only to the variance around the offset. The variance can be interpreted as the first order variation, whereas the offset as the zeroth order variation. In order to assess if the S/N of a telluric line was sufficient, we then only selected wavelengths for which the zeroth order variation was detectable.

Detecting the zeroth order variation is quite straightforward. To do so, the ratio time-series uncorrected from the telluric CCFs moments described in Sect.~\ref{sec:telluric} is shifted in the terrestrial rest-frame and the time-median of the time-series is taken to create a master telluric spectrum. A 1.5 interquartile sigma clipping was then performed to detect all the outliers which are proxies of the telluric locations. Such method was also used to form the wavelength set $\lambda_j$ on which to train the PCA for the ThAr bleeding (see Sect.~\ref{sec:thar}). %The telluric locations were extended of one resolution element in order to slightly extend the window around the telluric lines. 
The advantage of such data-driven method is that only the telluric lines with a high enough S/N will be selected. It improves the efficiency of the PCA which is not trained on noisy data.

\begin{figure}[h]
	\includegraphics[width=9cm]{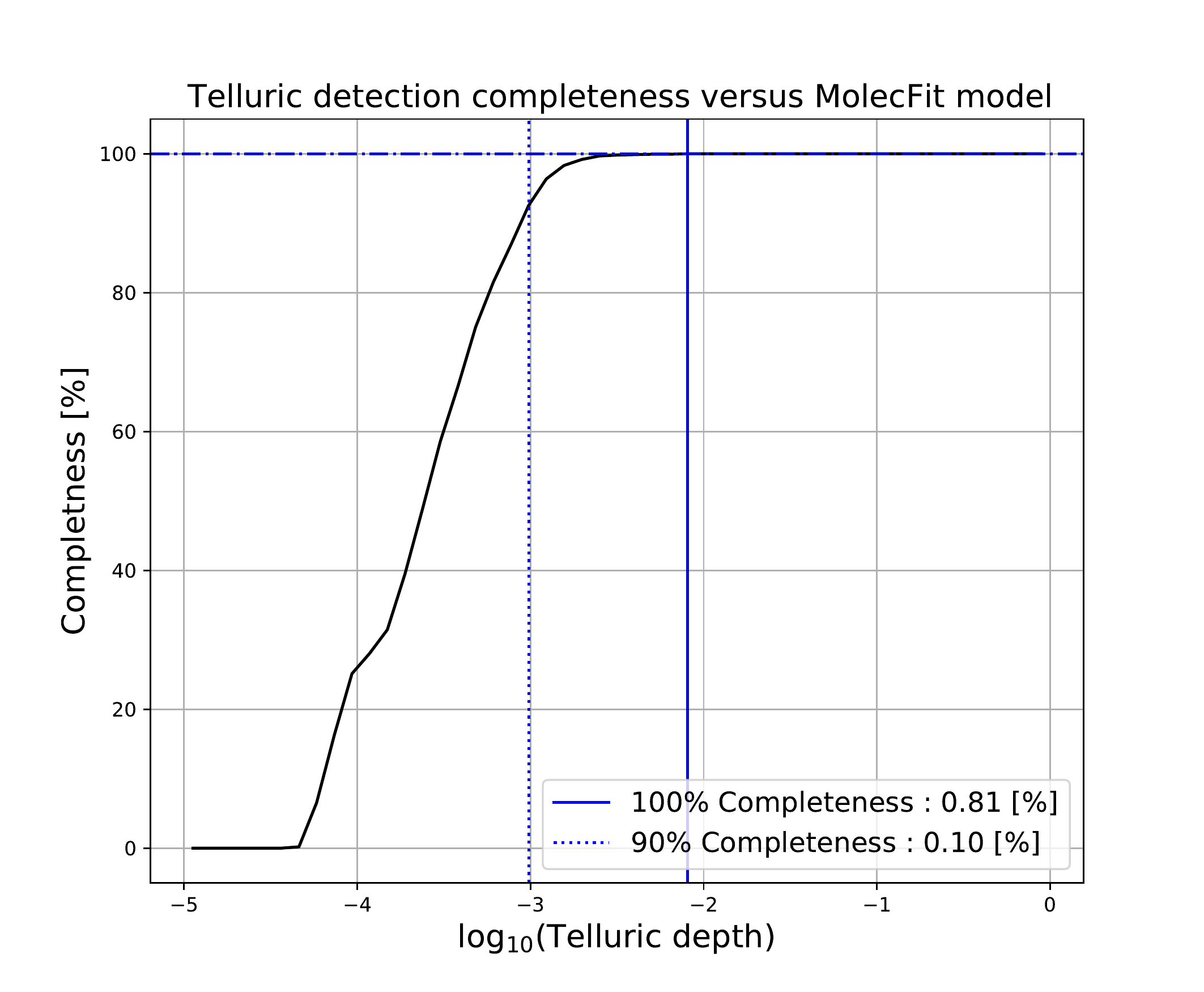}
	\caption{Computation of tellurics completeness of our detection algorithm compared to the MolecFit database for HD10700. We show here the cumulative distribution of the MolecFit tellurics lines undetected by our algorithm. We see here that all tellurics lines deeper than 0.81\% are detected. In addition, 90\% of the undetected telluric lines all present depth shallower than 0.10\%, which is coherent with the average noise dispersion around 0.21\% for HD10700 (see bottom panel Fig.~\ref{FigTelluric}).}
	\label{FigContamTelluric}
\end{figure}

Using the data of HD10700, we compared our telluric detection criterion with the MolecFit database in order to determine the completeness of our algorithm. To do so, we considered a MolecFit line as detected if this line was falling in a telluric region flagged by our pipeline. The cumulative distribution of the undetected telluric lines is displayed in Fig.\ref{FigContamTelluric}. As a comment, such analysis is a standard product of the pipeline, therefore allowing us to easily check if something wrong occurred and the expected level of correction reached, usually defined as the 90\% completeness level.

We see that all the lines deeper than 0.81\% are detected and 90\% of the undetected telluric lines are shallower than 0.10\% coherent with the noise level of the observations (0.21\% in the red, see bottom panel Fig.~\ref{FigTelluric}). For HD10180 and HD215152, the two other stars presented in the paper, the 90\% completeness level was situated respectively at 0.20\% and 0.25\%, still below the percent level.

\section{Ghost modelling and location extraction} \label{app:modeling}

In order to localise the ghosts on the detector, we first stacked two months of FLAT raw frames for fiber A and B, which allows to identify the ghosts produced by the science and the simultaneous reference fibers. We then masked the main echelle orders using the order localisation provided by the DRS. In the master raw frames obtained (see left panel of Fig.~\ref{FigGhost}), we can see several families of ghosts, that are characterized by their angle with respect to the main echelle orders.

To localise the exact position of the ghosts, we first measured their relative angle with respect to the main echelle orders. This was done by incrementally rotating the raw master image while interpolating it on the initial pixel grid and summing up the intensities of each column. By plotting the summed intensity as a function of rotational angle, clear peaks appear at certain angles, corresponding to the different families of ghosts. We detected three families of ghosts, one with a relative positive angle, and two others with relative negative angles. We only considered at this stage the family of ghost that was presenting the highest flux (the one with the negative relative angle closest to 0), as the other families were very dim and we could not see any impact of them in our river diagrams.
%\st{ are visible when a family is vertically aligned. The first family detected is at an angle of $0^\circ$ and is of course the physical order which are roughly perpendicular to the cross dispersion direction. We detected three others angles, one positive and two negatives. We only kept the first negative one being composed of the brightest ghosts family which can be detected by eye on the raw image. We dropped the others families since their flux intensities were not sufficient to properly fit their positions and no significant effects on river diagram was observed at their expected locations. Adding them to the train sample of the PCA was therefore reducing the goodness of the extracted components. 

\begin{figure}[h]
	\includegraphics[width=9cm]{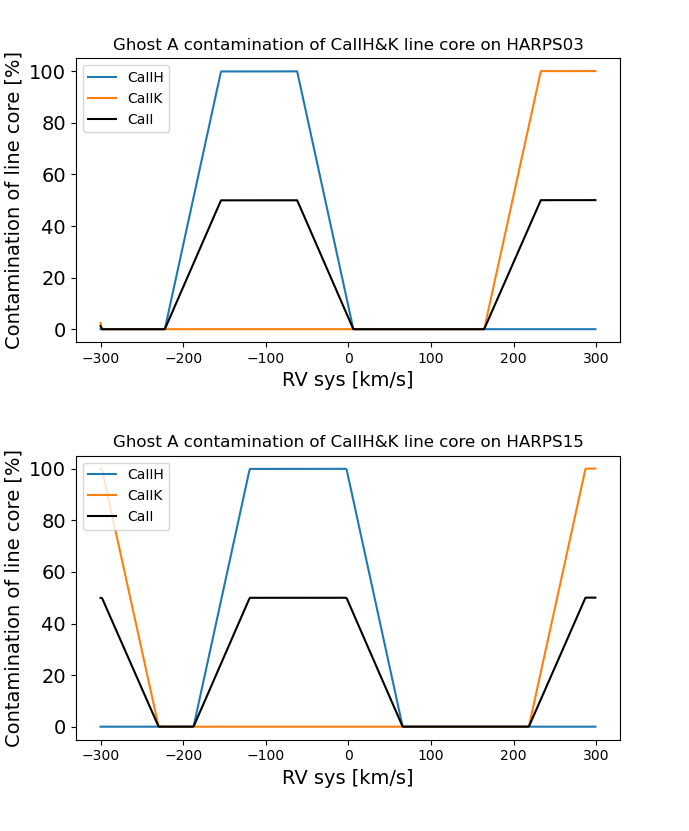}
	\caption{Computation of the covering fraction of ghosts from fiber A on the core of CaIIH\&K lines as a function of the systemic stellar velocity $RV_{sys}$. \textbf{Top:} HARPS before fiber upgrade. \textbf{Bottom:} HARPS after fiber upgrade.}
	\label{FigContamGhost}
\end{figure}

From the rotational angle and the position of the highest peaks, we can recover the position of the ghosts on the detector as straight lines. As ghosts do present a parabolic shape, we then looked for the maximum intensity around our linear approximation to obtain a localisation map for the ghosts. For each ghost, we extracted the flux at the localized position, and fitted a Gaussian profile. At the end of this process, we obtained the 2D map of ghost contamination seen in the middle panel of Fig.~\ref{FigGhost}. Finally, we selected the region where ghosts contaminate the science fiber by computing the intersection between the main orders and the ghost model (last panel Fig.~\ref{FigGhost}), and measured the relative intensity of ghosts by computing the ratio between the physical order and the ghost model intensities\footnote{We note that this relative intensity is the one computed for the spectrum of a Tungsten lamp, which has its own spectral energy distribution. It is likely that for a stellar spectrum, the relative intensity will be slightly different}. This process was applied on the FLAT master frames for fiber A and B in order to discriminate the ghosts originating from the science and the calibration fibers. 

Such static product can be used to measure the ghost contamination on the cores of the active CaII\,H\&K lines, where the contamination will solely depends on the stellar systemic velocity $RV_{sys}$ that will change the relative position between the lines' cores and the ghosts. On HARPS, only the ghosts from fiber A are sufficiently bright in that spectral region to be concerning. We extracted both chromospheric lines in short wavelength rectangular band-pass defined as $3933.66\pm0.45$ and $3968.47\pm0.45$ $\AA$ for the K and H lines respectively. By shifting those filters with different $RV_{sys}$ and computing the intersection with the ghost localisation, we can measure the covering fraction of ghosts on the cores of active lines as displayed in Fig.~\ref{FigContamGhost}. We see that during the full lifetime of HARPS, ghosts never contaminated simultaneously the core of both chromospheric lines. Note that such analysis is not considering the relative flux variation of ghosts and cores of the strong lines. Therefore, even a line completely covered by a ghost may present only a small relative contamination at the end if the CaII lines present a strong flux variation.

%\st{We created a simple metric by summing the intensities values in a zone of 10 pixels around the ghost localisation model, maximising the metric providing the best parabola curve. %Parabolic shape was then fitted by maximising the previous metric with an iterative process. The fit converge after a few iterations. 
%The second step consisted in measuring the ghost intensity profile which is at first order simply described by a Gaussian profile that can be fitted on the extracted intensities along the ghosts locations. We obtained by this process a 2D model of the ghost intensity on the detector (middle panel Fig.~\ref{FigGhost}). \st{The contamination were simply derived by computing the intersection between the physical order and the ghost model} (last panel Fig.~\ref{FigGhost}). \st{The relative intensity of the contamination was determined by computing the ratio between the physical order intensity and the ghost model intensity. As a remark, this relative intensity is the one computed from the Tungsten lamp which possesses its own spectral energy distribution different from a stellar energy distribution. }

\section{Continuum absorption correction} \label{app:continuum}

Our input products for YARARA are merged 1D spectra normalised by RASSINE, an alpha-shape algorithm valid under the assumption that the local maxima are probing the real continuum. As a consequence, an inaccurate continuum is often obtained for the blue part of the spectrum of G-K dwarfs or everywhere for M-dwarfs since this condition does not hold. Given the stellar parameters of a given star, it is possible to correct for continuum absorption using a stellar template. We used here the freely available POLLUX database \citep{Palacios(2010)} that contains normalised high-resolution spectra produced by MARCS \citep{Laverny(2012)} and ATLAS \citep{Kurucz(2005)} stellar atmosphere models. We note that MARCS models are 3D atmospheric models compared to 1D ATLAS models, but the ATLAS model grid of model is more extended and cover with a better resolution the $T$-$\log(g)$ space.

\begin{figure*}[h]
	\includegraphics[width=18.5cm]{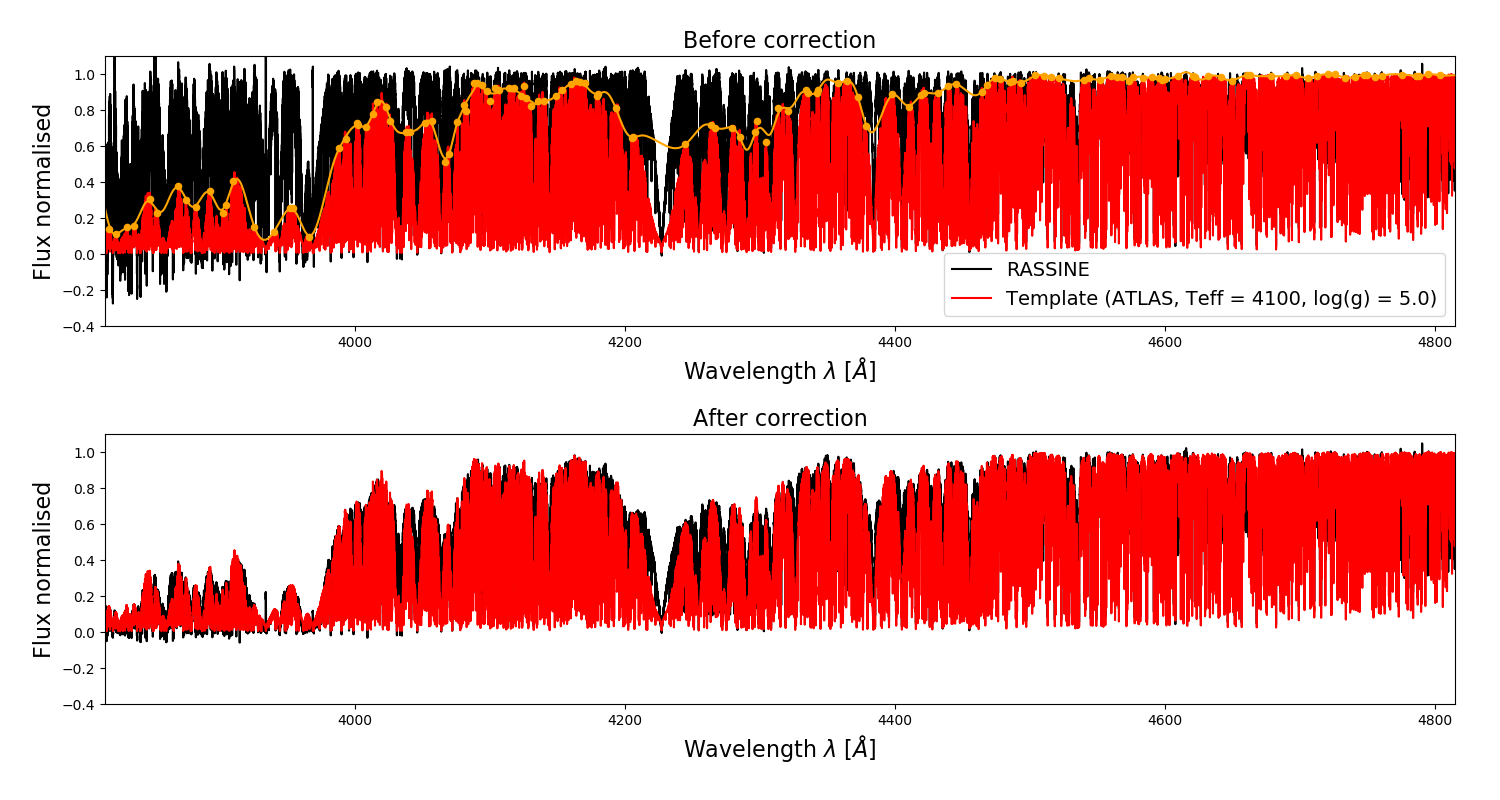}
	\caption{\textbf{Top:} Comparison between the GJ576 master spectrum coming out from RASSINE (black curve) and the best-fitted template model (red curve). The star is a M1.5 dwarf for which the best stellar template in ATLAS is a model with $T=4100$K and $\log(g)=5.0$. We show in orange, the correction curve for the continuum absorption. \textbf{Bottom:} RASSINE master spectrum after template correction. The accuracy in the blue is considerably improved.}
	\label{FigContiGJ654}
\end{figure*}

We show in Fig.~\ref{FigContiGJ654} an example of continuum absorption correction for GJ576, a M1.5 dwarfs.
The first step consists in getting an approximate value for the effective temperature and surface gravity of GJ576. This is done by minimizing the $\chi^2$ between the model and three strong lines in our RASSINE master spectrum, namely $H_{\alpha}$, the Na doublet and the MgI triplet since the broad wings of those lines are known to be pressure and temperature sensitive \citep{Gray(2005)}. We note that before minimizing the $\chi^2$ between different models, the master spectrum is shifted in the stellar rest frame and the stellar templates are interpolated on the same wavelength grid as the master. We then performed a rolling 90-percentile with a window of 10 $\ang$ on the stellar template and kept only the points above the obtained curve. Finally, only the closest maxima to the RASSINE anchors points are kept and a cubic interpolation is performed between these local maxima before applying a Savitchy-Golay filtering with a window of 10 $\ang$. The obtained continuum (see orange curve in Fig.~\ref{FigContiGJ654}) is used to divide the RASSINE master spectrum (black curve top panel), which allows to correct for the continuum absorption, as we can see in the bottom panel.

\end{appendix}

\end{document}